\documentclass[manuscript,screen,noacm]{acmart}
\usepackage{longtable}
\AtBeginDocument{%
  \providecommand\BibTeX{{%
    \normalfont B\kern-0.5em{\scshape i\kern-0.25em b}\kern-0.8em\TeX}}}

\setcopyright{none}
\renewcommand\footnotetextcopyrightpermission[1]{} 
\begin{document}

\title{User Attitudes to Content Moderation in Web Search}

\author{Aleksandra Urman}
\affiliation{%
  \institution{University of Zurich}
  \country{Switzerland}}
\email{urman@ifi.uzh.ch}

\author{Aniko Hannak}
\affiliation{%
  \institution{University of Zurich}
  \country{Switzerland}}
\email{hannak@ifi.uzh.ch}

\author{Mykola Makhortykh}
\affiliation{%
  \institution{University of Bern}
  \country{Switzerland}}
\email{mykola.makhortykh@unibe.ch}

\renewcommand{\shortauthors}{Urman et al.}

\begin{abstract}
  Internet users highly rely on and trust web search engines, such as Google, to find relevant information online. However, scholars have documented numerous biases and inaccuracies in search outputs. To improve the quality of search results, search engines employ various content moderation practices such as interface elements informing users about potentially dangerous websites and algorithmic mechanisms for downgrading or removing low-quality search results. While the reliance of the public on web search engines and their use of moderation practices is well-established, user attitudes towards these practices have not yet been explored in detail. To address this gap, we first conducted an overview of content moderation practices used by search engines, and then surveyed a representative sample of the US adult population (N=398) to examine the levels of support for different moderation practices applied to potentially misleading and/or potentially offensive content in web search. We also analyzed the relationship between user characteristics and their support for specific moderation practices. We find that the most supported practice is informing users about potentially misleading or offensive content, and the least supported one is the complete removal of search results. More conservative users and users with lower levels of trust in web search results are more likely to be against content moderation in web search.
 
\end{abstract}




\maketitle

\section{Introduction}
The amount of information available online nowadays necessitates the use of web search engines (SEs) that filter and rank information in response to user queries. Internet users turn to SEs on a daily basis and put high trust in the information they find through web search \cite{schultheis_we_2018,urman_you_2021}. At the same time, while SEs are often perceived as impartial mechanisms for information retrieval \cite{tripodi_propagandists_2022}, scholars have documented numerous biases and inaccuracies in web search outputs over the years (e.g., \cite{kay_unequal_2015,noble_algorithms_2018,zhang_quality_2015}). Others have highlighted the differences across SEs and their localized outputs
in the prevalence of low-quality content such as materials promoting conspiracy theories \cite{urman_where_2022} or the availability of crucial information such as suicide helpline numbers 
\cite{scherr_algorithms_2022}. The observed discrepancies partially stem from the differences in the search algorithms employed by different SEs and the availability of certain content in different languages and can potentially in part be attributed to the ways content moderation is implemented for individual SEs. 

In scholarly research, content moderation (CM) is discussed primarily in the context of social media but other online platforms, including SEs, also employ it - though in a highly intransparent manner \cite{gillespie_custodians_2018,gorwa_algorithmic_2020,urman_how_2023}. The official documents of the most popular search engines confirm this as they outline how SEs utilize the practices of either informing users about potentially dangerous websites, downgrading low-quality outputs or removing them altogether \cite{google_general_2022,microsoft_how_nodate,yandex_how_nodate,yandex_signs_nodate,duckduckgo_news_nodate,google_information_2020}. Importantly, generally within this paper - for instance, when describing search companies' moderation practices - we understand content moderation broadly, similarly to  \cite{gorwa_algorithmic_2020}. That is, we discuss CM including the moderation of content that is illegal, not just content that the search companies themselves regard as necessary to moderate. However, our examination of \textit{user attitudes to CM} does not concern illegal content since the types of content deemed illegal and forms of its moderation - i.e., its removal - are not determined by search companies and are outside their control. Thus, when it comes to illegal content, user attitudes to this specific form of moderation are less consequential, and arguably need to be explored in relation to the users' perceptions of relevant laws in their countries, not search companies' policies and practices. 

Content moderation on online platforms becomes an increasingly salient political issue, at least in the Western democracies \cite{alizadeh_content_2022} since its implementation can directly affect users' access to information and thus socio-political processes. At the same time, user support for content moderation is imperative for its successful implementation. For these reasons, numerous studies have examined the determinants of support for content moderation online. However, to date, this, to the best of our knowledge, has been examined only in the context of social media, and despite the high reliance of the public on SEs and the active use of moderation by search engines, moderation practices in web search have not been systematized and user attitudes to them have not been examined. Since SEs and social media are distinctly different types of platforms used by different groups of users and for different purposes, we believe that the findings on content moderation from social media domain do not necessarily translate directly into the SE domain. Thus, the lack of research on user perceptions of content moderation in web search specifically constitutes a clear research gap that we aim to address with the present study.

We use the data from a survey of a demographically representative sample of the US adult population (N=398) to examine the levels of user support for different content moderation practices in web search in relation to potentially misleading and potentially offensive content. We analyze which user characteristics and opinions such as demographics, ideology, or trust in SEs are associated with higher/lower support for specific moderation practices. In order to construct our survey questions in a way that covers actual moderation practices that are currently in use by search engines, we first systematize these practices based on the search companies' official documents and media statements. As such systematization has not been done before, to the best of our knowledge, we suggest that the resulting overview is a contribution on its own. We hope it will be helpful for other scholars examining user interactions and information quality in web search as well as content moderation across different types of online platforms. We present this overview preceding the study design. We also discuss our findings juxtaposing them against the actual CM practices of SEs and the findings on the relationships between user characteristics and support for content moderation previously documented by scholars in the context of social media.

In the next sections, we first outline relevant observations from the previous work on the usage of SEs and the quality of search outputs. Then, we present an overview of content moderation practices in web search and shortly systematize them. This is followed by an overview of related work on user attitudes to CM in the context of other types of online platforms such as social media. After that, we detail specific Research Questions and Hypotheses building on the related work and the systematization of content moderation practices presented in the previous sections. Finally, we outline the methodology, describe and discuss our results.

\section{Related work on web search usage and quality of search outputs}
Individuals regularly use search engines to gather information on a variety of topics and facilitate navigation through contemporary high-choice media environments \cite{urman_you_2021}. The fact that Google - the biggest SE by market share - is one of the most frequented websites worldwide further highlights how much people rely on web search in their daily lives. Further, not only do people regularly use SEs, they also trust their outputs as much as the information from journalistic media \cite{edelman_2021_2021}. This is not a recent phenomenon - high trust in search outputs has been consistently observed by scholars for over a decade \cite{hargittai_trust_2010,pan_google_2007,schultheis_we_2018}. Together with the increasing abundance of online information that is almost impossible to navigate without SEs, this high trust turns SEs into major information gate-keepers.

While trust in search outputs is high and has remained stable over time, numerous studies showed that search results are prone to inaccuracies and biases. For example, research has demonstrated that SE outputs exhibit different forms of gender and/or racial bias in search results about specific social groups \cite{kay_unequal_2015,noble_algorithms_2018,ulloa_representativeness_2022,urman_foreign_2022,metaxa_image_2021}. Recent scholarship also shows that exposure to such stereotyped or biased representations of people via SEs can increase people's prejudices against the groups portrayed in a biased manner \cite{vlasceanu_propagation_2022}. 

One domain where the prevalence of misleading information in web search results is particularly concerning, and thus has attracted a lot of scholarly attention, is public health. While the share of inaccurate or low-quality outputs varies by specific health domain, scholars highlight that overall, the quality of health-related outputs remains problematic (see \cite{zhang_quality_2015} for a systematic literature review prior to 2015 or \cite{10.1145/3079452.3079483,info:doi/10.2196/18444} for more recent evidence). In domains other than health, recent comparative studies show that the prevalence of low-quality information such as results promoting conspiracy theories or distorting historical facts differs drastically by SE and the language in which the search is performed \cite{makhortykh_hey_2021,makhortykh_memory_2022,urman_where_2022}. The language-based differences in the quality of search results specifically on Google are further documented by a number of other recent studies \cite{toepfl_who_2022,arendt_investigating_2020,scherr_algorithms_2022,scherr_equal_2019}. Such cross-engine and cross-language differences can, in turn, contribute to digital divides between users \cite{scherr_algorithms_2022}. The documented differences are likely attributed to the differences in the availability of specific sources across languages and discrepancies in web search algorithms. However, it is possible that some of the differences in the share of low-quality (e.g., misleading, conspiratorial, or offensive) content have to do with the differences in the content moderation practices of SE companies across languages and contexts. 

Content moderation in web search is especially crucial given users' high trust in and reliance on search outputs as well as a common belief that search engines present "unbiased" information \cite{tripodi_searching_nodate}. Relevant research provides evidence that search results can affect individual opinions or (perceived) knowledge \cite{epstein_search_2015,vlasceanu_propagation_2022,fisher_searching_2015,xu_how_2021,knobloch-westerwick_confirmation_2015}. Hence, low-quality content can effectively contribute to the spread of misinformation and the propagation of harmful stereotypes, and thus arguably needs to be moderated. On the other hand, there exists a risk of overmoderation or the abuse of content moderation practices resulting in de-facto censorship of certain search results as is the case in some authoritarian regimes that have tight control over local search engines \cite{makhortykh_story_2022}. In the next section, we provide an overview of the state of content moderation across web search engines.

\section{Overview and systematization of content moderation practices in web search}
SEs formally fit the criteria commonly used to define online platforms \cite{gillespie_custodians_2018}: they host and organize users' content without having produced or commissioned that content and their infrastructure enables organization and distribution of information, including for-profit uses of user data (e.g., for advertising). Another common criterion used to define platforms is: "platforms do, and must, moderate the content and activity of users using some logics of detection, review, and enforcement" \cite{gillespie_custodians_2018}. 

In the case of SEs, content moderation (CM) practices can take different forms. One of them relates to the prioritization of specific types of information sources. Today's search engine outputs are typically structured in the form of vertically organized lists. This contributes to the users' likelihood to perceive top results as more important or reliable \cite{pan_google_2007,tripodi_searching_nodate}, and to click on top results more often \cite{pan_google_2007,urman_you_2021}. There is evidence that presenting search results in a different form - e.g., as a tabular "grid" rather than a list, - mitigates these tendencies and leads to users searching in a more focused manner \cite{kammerer_effects_2013}. Thus, the decision to organize outputs as lists itself affects user behavior. The list-based organization of information increases the importance of the way search results are \textit{ranked}. It is not only important \textit{which} results are displayed in response to a search query, but \textit{how} - i.e., in what order, - they are displayed. Thus, in web search results not only removal but downgrading of certain outputs - and thus the reduction of their visibility \cite{gillespie_not_2022} - is a highly viable moderation practice that many SEs actively employ.

In contrast to the substantive volume of scholarship on social media content moderation \cite{gillespie_content_2020,gillespie_not_2022,gerrard_beyond_2018,ganesh_countering_2020,riedl_antecedents_2022,morrow_emerging_2022,myers_west_censored_2018}, web search content moderation remains a rather under-studied subject. We have not been able to find empirical studies examining the ways different moderation practices work across SEs or the ways they are implemented. Hence, we provide some background on web search moderation based on the information from the documentation and statements by SEs representatives. To infer whether and how SEs moderate their outputs, we have checked the statements made by the companies in the official documentation and the claims coming from their official representatives  - e.g., through social media and news media comments. Our analysis here is limited, and we provide only more general information since the detailed examination of related documents and statements arguably merits a standalone paper and is out of the scope of the present study. 

 Importantly, we focus on the general moderation practices and do not cover anything specific to the so-called SafeSearch mode that is implemented by some engines. Further, our overview originally corresponded to the practices employed by SEs in the second half of 2022 - to align with the time when the survey for our study was conducted. As such practices and policies change overtime, we revisited this section in September 2023 when preparing the final version of the paper, and have documented the observed changes (or lack thereof) in the companies' policies and practices.
 
 We focus on the major SEs by market share in the US \cite{statcounter_search_2022} since our study is US-focused. Notably, the same engines are the most popular ones in most Western countries. This includes Google, Bing, Yahoo!, DuckDuckGo, Yandex, and Ecosia according to \cite{statcounter_search_2022}.

\subsection{Content moderation on Google}
Google has published a White Paper on the way it moderates content across its services \cite{google_information_2020}. This includes not only web search but also other services such as Google Maps (with a bulk of the report devoted to YouTube). Among the actions Google takes to limit the spread of harmful or misleading content are removals and reduction of exposure to it (e.g., through not recommending such content). It is unclear how these are applied in web search. 

It is known that "quality" is one of the characteristics taken into account by Google when ranking content. The operationalization of quality, however, is ambiguous. Google employs ~14000 (as of 2022 \cite{google_search_2022}) "Quality Raters" across the world that rate different aspects of web pages resurfacing in search results, including whether these pages are potentially harmful - e.g., offensive or containing misinformation (see detailed guidelines and definitions from Google as of 2022 \cite{google_general_2022}; also see \cite{meisner_labor_2022} for more details on the work of Quality Raters). At the same time, it is ambiguous how these ratings impact the ranking 
of pages deemed harmful in search results. Google simply states\footnote{The statement was originally accessed in 2022, and was still available in the same form in September 2023.} "We work with external Search Quality Raters to measure the quality of Search results on an ongoing basis. Raters assess how well content fulfills a search request, and evaluate the quality of results based on the expertise, authoritativeness, and trustworthiness of the content. These ratings do not directly impact ranking, but they do help us benchmark the quality of our results and make sure these meet a high bar all around the world." \cite{google_rigorous_2022}. The hidden labor of Quality Raters is entangled with the different ideological and economic layers of the algorithmic development \cite{bilic_search_2016}, and it remains unclear how this affects the actual composition of search results.

Additionally, Google states that it attaches warning notes to website links that can be potentially dangerous for the users and their computers - i.e., those suspected of phishing or spreading malware \cite{google_manage_2022}.

\subsection{Content moderation on Yahoo!}
On Yahoo! the implementation of content moderation
is even more opaque than on Google based on the company's official documents. For instance, in a FAQ page on search result removal Yahoo! states that it has no control over what is published outside of its network \cite{yahoo_remove_2022}. However, there is a note that if users' personal information is published, they can seek assistance from Yahoo! to remove the website publishing such information from search results \cite{yahoo_remove_2022}. In addition, the company's description of its international Search Services privacy practices includes a statement that "Users who are European residents can request that certain URLs be blocked from search results in certain circumstances." \cite{yahoo_search_2022}. The specific circumstances however are not specified.

Based on this information, it can be implied that Yahoo! sometimes removes search results (e.g., when it comes to illegal content or personal information), but it is unclear 
how such decisions take place and whether the search engine additionally removes or downgrades any links containing  misinformation or offensive content. We did not find any updates on this in Yahoo!'s documentation as of September 2023.

\subsection{Content moderation on Bing}
Microsoft, the owner of Bing, as of 2022 clearly stated that it removes search results under certain circumstances which include, for example, government requests or requests from companies/individuals when it comes to content that is illegal - e.g., content dealing with child abuse or copyright infringing - or in the cases of spam \cite{microsoft_how_nodate}. The company also noted that when it removes content, it mentions this at the bottom of the search results page \cite{microsoft_how_nodate}. In 2022, we did not find information on the downranking of search results, we did find a statement from Microsoft that in some cases instead of removing a result, the company accompanies it with a warning to the users - e.g., for the websites that potentially contain malware or sell illegal pharmaceuticals \cite{microsoft_how_nodate}. How exactly the decisions on the addition of warnings or content removal are made is unclear. In September 2023, the information provided by Microsoft regarding content moderation was slightly different than that we originally read in 2022. Specifically, the company has now added mentions of downranking as a form of content moderation "where the content violates local law, or Microsoft’s policies or core values" \cite{microsoft_how_nodate}. The company as of September 2023 mentions it strives for such actions to be "narrowly tailored" \cite{microsoft_how_nodate}, however, how exactly such decisions are made is still not clarified.

\subsection{Content moderation on DuckDuckGo}
It is unclear whether and how DuckDuckGo moderated search results up to 2022. Several analyses in 2021 found that DuckDuckGo outputs often promote conspiratorial content \cite{thompson_fed_2022,urman_where_2022}. However, shortly after Russia invaded Ukraine in February 2022 DuckDuckGo's CEO and founder, Gabriel Weinberg, announced that the search engine has been "rolling out search updates that down-rank sites associated with Russian disinformation" \cite{gabriel_weinberg_yegg_like_2022}. DuckDuckGo's official webpage at the time of writing also states that low-quality news media are downgraded in the search results - albeit users should still be able to find the links to them as only illegal content is completely removed \cite{duckduckgo_news_nodate}. The company also clarifies that in their assessment of the quality of news media they "rely on multiple non-governmental and non-political organizations that specialize in objectively assessing journalistic standards. To take any ranking action using this factor, we must see at least three of these organizations independently assess a site as having extremely low journalistic standards and also see that none of these organizations has assessed the same site as having even somewhat robust journalistic standards" \cite{duckduckgo_news_nodate}. It is unclear, however, in which countries these organizations function and whether this applies only to the US-based and/or English-speaking media or those in other languages and/or other parts of the world. As of September 2023, the company has added an additional explanation about its moderation processes in the section about "common misconceptions" regarding DuckDuckGo. Specifically, DuckDuckGo, in connection to the potential censorship of search results, states "Our search ranking is strictly non-political, meaning we don’t evaluate or otherwise take into account any potential political bias or leanings of websites in our search result rankings." \cite{duckduckgo_does_nodate}. Additionally, on a page devoted to a misconception about Russian search results, the company states "We also do not evaluate the “truth” of any particular news story or narrative." \cite{duckduckgo_did_2023}. The latter is a notable distinction between DuckDuckGo's policies and that of other engines such as Google that state they provide warnings with regard to misleading content - and thus implicitly evaluate the "truth" of different sites and narratives.

\subsection{Content moderation on Ecosia}
We could not find information on content moderation on Ecosia in the SE's official documents and statements or news reports neither in 2022 nor in 2023.

\subsection{Content moderation on Yandex}
Yandex states that for certain violations of its policies, it might remove a link from search results completely, demote it in results and/or also accompany it with a warning to the users - e.g., that a website might be potentially dangerous \cite{yandex_how_nodate,yandex_signs_nodate}. The decision depends on the type of policy violation with the correspondence between demotion/deletion/warning and violation types clearly outlined \cite{yandex_how_nodate,yandex_signs_nodate}. We found the same was true as of September 2023. In a way, Yandex is more transparent than other SEs about the content moderation practices it employs. At the same time, it is a Russian search engine, and according to reports, it sometimes removes or alters content in ways that favor the Russian government 
\cite{makhortykh_story_2022,lomas_russian_2022}.

\subsection{Summary}
Overall, the content moderation policies of the most popular SEs are rather opaque. At the same time, we can systematize the information about existing practices and derive 3 main types of CM practices that are currently used by the SEs:
\begin{itemize}
    \item \textbf{Informing users} - for instance, through adding "warning labels" to certain types of content such as misleading content. We found confirmations that this is done by Google, Bing and Yandex, according to their official statements \cite{yandex_how_nodate,yandex_signs_nodate,microsoft_how_nodate,google_general_2022}.
    \item \textbf{Reducing the reach of certain content} - mostly through downgrading it in search results. This practice is explicitly mentioned by DuckDuckGo, Google and Yandex \cite{yandex_how_nodate,yandex_signs_nodate,duckduckgo_news_nodate,google_information_2020}.
    \item \textbf{Removing certain content} - in certain cases, SEs remove content from search results altogether. This practice is confirmed to be used by Google, Bing, DuckDuckGo, Yahoo and Yandex \cite{google_information_2020,microsoft_how_nodate,yahoo_remove_2022,yandex_how_nodate,yandex_signs_nodate,duckduckgo_news_nodate}. Most often, based on what we inferred from the cited companies' documents and statements, removals take place in the cases when the indexed content violates local laws.
\end{itemize}

In addition, we observe that at least according to the companies' official statements, SEs currently focus on moderating two main types of content: illegal content and misleading 
content. This is in contrast to other platforms (e.g., social media) which typically also moderate offensive content such as hate speech \cite{gillespie_custodians_2018}. It is unclear what drives the difference between SEs and social media in this regard - the difference in the nature of the platforms, company cultures, or perceived user expectations.

The observations on the SEs' content moderation practices outlined in this section inform our research questions (RQs) as detailed 
below.

\section{Related work on user attitudes towards content moderation}
While all online platforms including SEs moderate content in one way or another \cite{gillespie_custodians_2018}, thus directly influencing information exposure and experiences of their users, the user attitudes towards content moderation practices so far have been explored only to a limited extent and, to the best of our knowledge, exclusively for social media platforms. There is thus a clear research gap with regard to the user attitudes towards content moderation practices in web search that we aim to address. Before outlining concrete RQs and hypotheses in the next section, we first present a summary of the findings on attitudes towards content moderation on social media as they inform our research.

A 2019 survey by YouGov showed that around 45\% of respondents in the US support the idea of CM by social media in general \cite{ballard_most_2019}. The same survey however also demonstrated the drastic differences in the attitudes to content moderation between liberals and conservatives with the former being more likely to support content moderation than the latter \cite{ballard_most_2019}. A similar observation was made in a different study from 2022 \cite{kozyreva_free_2022}. Another study conducted in the US did not find a relationship between political partisanship and support for CM but found that age and level of education are significantly related to CM support with older and higher-educated users more likely to be in favor of it \cite{riedl_antecedents_2022}. Yet another analysis conducted in the US has shown that there is bipartisan support for labeling certain content (i.e., informing users) as a form of content moderation \cite{wihbey_bipartisan_2021}. In one study, sex and race of the respondents were not associated with attitudes towards CM \cite{riedl_antecedents_2022}. At the same time, a survey among the US youth found that young women were more likely than young men to support CM \cite{schoenebeck_youth_2021}. Other analyses on the topic - some of which relied on in-depth interviews rather than surveys - have concluded that opposition to content moderation often is connected to the users' low trust in the companies' ability to make fair and transparent moderation decisions and/or beliefs that CM processes and outcomes are biased in a certain way (e.g., affected by political or business interests) \cite{jhaver_did_2019,duffy_platform_2022,myers_west_censored_2018,saltz_encounters_2021}. Another factor that previous research has found to be associated with lower/higher support for content moderation and specific moderation decisions in relation to offensive content specifically is the exact wording used in a social media post that is to be moderated \cite{pradel_users_2022}. 

Additionally, researchers have found that users' attitudes to content moderation differ, depending on who - or what, in the case of algorithms - makes a decision to moderate certain content. For instance, an experimental study of Facebook users found that the participants perceived moderation decisions taken by expert panels as more legitimate than those taken by the algorithms or juries \cite{pan_comparing_2022}. Further, \cite{ozanne_shall_2022} established that social media users have less trust in moderation decisions that are coming from AI, as compared to when the moderation decision is taken by a human or when the moderation source is ambiguous. A similar observation was described by \cite{calleberg_making_2021}. In addition, experimental research has shown that users' perceptions of fairness and accountability in the context of CM decisions taken by the algorithms are not influenced by the presence of the right to appeal, regardless of the appeal formats tested by the researchers \cite{vaccaro_at_2020}. At the same time, users' levels of trust in moderation decisions taken by the algorithms vs humans are related to their ideological orientation - e.g., researchers established that conservatives in the US are more likely to trust moderation decisions when they are taken by AI rather than humans \cite{molina_does_2022}, once again highlighting the relation of ideology to the users' attitudes towards content moderation. 

As this overview demonstrates, there is a lot of conflicting evidence regarding user attitudes toward content moderation in the context of social media platforms. Despite the apparent contradictions, however, several patterns emerge: user demographics, political opinions, and trust in the platforms tend to be associated with the users' support for CM (on social media). We rely on these findings in formulating our research questions and hypotheses.

\section{Research questions and hypotheses}
In the previous sections, we have shown that 1) SEs are highly trusted and relied on by the users for retrieving correct and "unbiased" information, yet there is consistent evidence of biases and inaccuracies being present in web search results and varying across SEs; 2) SEs engage in diverse forms of content moderation - informing users, reducing the reach of content or removing content - in relation to illegal or false content, but not to offensive content despite it being moderated by other types of online platforms; 3) there is no evidence regarding user preferences on CM in web search, but research about user attitudes towards CM on other platforms shows that these attitudes are influenced by demographic characteristics, political opinions and trust in platforms. Based on this, we formulate specific RQs and hypotheses to address the existing research gap with regard to user attitudes to content moderation in web search.

For the first RQ, we aim to examine general user attitudes towards different forms of content moderation in web search. Here and in other RQs we focus on two specific types of content that might be subject to moderation: misleading/false content and potentially offensive content. This is informed by the fact that these two types of content are currently moderated by online platforms such as social media (in addition to content that is explicitly illegal) but only one of them (i.e., false content) seems to be moderated by SEs. Answering our RQs will allow us to establish whether this divergence in moderation practices corresponds to the user expectations. 

While we formulate the RQs below in general terms, we in fact examine user preferences for CM and their relation to user demographics and opinions with a breakdown of user preferences for three distinct practices employed by SEs as identified in the previous sections: informing users; reducing the reach of content; removing content. Hence, we examine user preferences separately for each of these practices of moderating misleading or offensive content. Importantly, we note that we interpret the reduction of the reach of specific content here only as a moderation practice. The reach of certain content would always be reduced (or, conversely, amplified) by search engines as they rank search outputs. However, we do not interpret the reduction of reach of some content in this case as moderation. We treat the reduction of reach as a form of moderation \cite{gillespie_not_2022} when a company specifically configures its algorithm to downrank certain sites in search output due to the nature of the content there, as compared to other websites that do not contain the content of that type (e.g., offensive or misleading). 

The RQs are formulated as follows:

\begin{itemize}
  \item \textbf{RQ1:} What are users' preferences on content moderation in web search?

  This RQ is divided into two sub-RQs corresponding to two different types of content: false/misleading content and content which some users might find offensive.
  \begin{itemize}
      \item \textbf{RQ1a}: What are users' preferences on content moderation in web search in relation to \textit{misleading or false content}?
      \item \textbf{RQ1b}: What are users' preferences on content moderation in web search in relation to \textit{potentially offensive content}?
  \end{itemize}
\end{itemize}

In RQs 2 and 3 we go beyond the descriptive analysis of CM preferences and evaluate how these preferences relate to different user characteristics. Within RQ2 we focus on misleading/false content; within RQ3 we focus on potentially offensive content. 
\begin{itemize}
  \item \textbf{RQ2:} How do user preferences for the moderation of \textit{misleading or false content} in web search relate to user characteristics? 
  \item \textbf{RQ3:} How do user preferences for the moderation of \textit{potentially offensive content} in web search relate to user characteristics? 
\end{itemize}

The two RQs are divided into sub-questions focused on specific user characteristics. Specific characteristics we choose to examine as being potentially relevant for CM preferences are informed by the prior research on user support for CM on other types of platforms and include user demographics (age, sex, race, level of education), political leaning (on the left-right spectrum), trust in the platforms and the perceived independence of the platforms. Additionally, motivated by the findings that the prevalence of biased and/or false information differs drastically across SEs, we also examine how the use of specific SEs is related to CM support. Since the examined characteristics and opinions are the same for both RQ2 and RQ3, we list dedicated sub-RQs only once (e.g., as RQ2/3a, RQ2/3b, etc).
\begin{itemize}
  \item \textbf{RQ2/3a}: How do user preferences for content moderation in web search relate to users' \textit{demographic characteristics (age, sex, race, level of education)}?
  \item \textbf{RQ2/3b}: How do user preferences for content moderation in web search relate to users' \textit{political (left-right) leaning}?
  \item \textbf{RQ2/3c}: How do user preferences for content moderation in web search relate to users' \textit{trust in web search}?
  \item \textbf{RQ2/3d}: How do user preferences for content moderation in web search relate to users' \textit{frequency of use of specific search engines}?
  \item \textbf{RQ2/3e}: How do user preferences for content moderation in web search relate to users' \textit{assessments of web search platforms' independence from undue political and business interests}?
\end{itemize}

As findings on the relationship between users' demographic characteristics or political opinions and support for CM on other platforms are contradictory, we do not formulate hypotheses in relation to this relationship and rather aim to explore the potential relationships in the context of web search. However, since previous research consistently shows that trust in platforms is related to the users' likelihood to support platforms' CM practices, while a belief that CM practices are biased due to political or business interests is related to lower support for CM, we hypothesize that the same effects will be present in the context of web search and formulate the following hypotheses connected to RQ2/3c and RQ2/3e:
\begin{itemize}
  \item \textbf{H1}: Users with higher levels of trust in SEs will be more likely to be in favor of CM in web search.
   \item \textbf{H2}: Users with higher levels of confidence in SE's independence will be more likely to be in favor of CM in web search.
\end{itemize}

\section{Methodology}
To address the research questions outlined above, we conducted a survey of a representative (in terms of age, sex, ethnicity) sample (N=398) of the US adult population, administered through Qualtrics and recruited through Prolific using the platform’s representative sampling functionality (see \cite{prolific_representative_2022}). We chose to focus on the US as it is a democratic country with a high internet penetration rate; further, most of the research on CM-related attitudes on other types of platforms (social media) so far focused on the US (e.g., \cite{kozyreva_free_2022,riedl_antecedents_2022,ballard_most_2019}), thus conducting analysis in the US enables us to connect our findings to those from other platforms. All responses were collected on August 22, 2022. In our sample, 50.2\% of respondents were female; mean age = 45.87, median = 46; 13.57\% of respondents were 18-25 years old (y.o.), 18.84\% - 26-35 y.o.;  32.91\% - 36-55 y.o.; 21.61\% - 56-65 y.o.; 13.07\% - 65+ y.o.; 76.13\% self-reported to be White, 12.81\% Black, 5.79\% Asian, 2.51\% Mixed,  2.76\% Other.

\subsection{RQ1}
To measure user attitudes towards different web search content moderation practices and thus address RQ1, we have adapted survey items used in \cite{atreja_what_2022} in the context of social media. For the content moderation practices in relation to misleading information, we used the following question:

"Some websites on the internet contain misleading content. When it comes to displaying links to such sites, search engines can take one of the following actions:

\begin{itemize}
    \item Inform users. For example, by showing a “misleading” icon next to the link to a misleading site in web search results.
    \item Reduce the audience that can see links to misleading websites without removing them. For example, by showing the link only on the second or third page of search results but not on the first page.
    \item Remove links to misleading websites from search results.
\end{itemize}

How much do you personally support or oppose taking any of these actions when it comes to websites with misleading content?"

\begin{figure}[h]
  \centering
  \includegraphics[width=\linewidth]{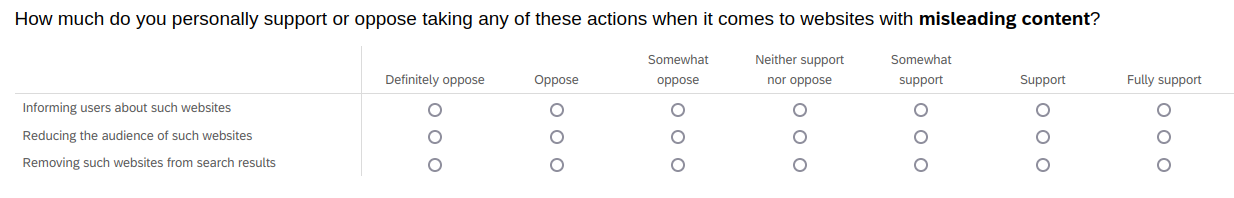}
  \caption{Survey response matrix for survey item on content moderation of misleading content.}
  \label{fig:surveyscale}
\end{figure}

Then, the respondents were presented with a response matrix where they could mark their level of support for each of the measures on a 7-point Likert scale (see example in Fig. \ref{fig:surveyscale}).

To measure the participants' attitudes to content moderation of potentially offensive content, we used a similarly formulated question followed by a response matrix similar to that in Fig.\ref{fig:surveyscale}. The difference here was that instead of the term "misleading content" in this case we used "content that some users can find offensive or disturbing". 

The responses to the questions on CM practices were used to calculate descriptive statistics necessary to answer RQ1a, RQ1b. In addition, to establish whether the discrepancies in the levels of support towards different types of content moderation observed through descriptive analysis are statistically significant, we performed a Kruskall-Wallis rank sum test followed by pairwise comparisons using Wilcoxon signed rank test with Bonferroni adjustment to control for multiple comparisons \cite{r_core_team_stats-package_2023}. We opted for these tests instead of, e.g., MANCOVA, as our variables are ordinal in nature and not normally distributed. Thus, MANCOVA assumptions would have been violated \cite{french_multivariate_nodate}, and the chosen tests are suitable for our data.

\subsection{RQs 2, 3, Hypotheses 1, 2}
To answer RQ2 and RQ3 with all the corresponding sub-RQs as well as test hypotheses H1 and H2, we used regression analysis. Specifically, we ran ordinal logistic regression models using the 6 variables on the attitudes to CM practices (3 for each practice in relation to misleading content and 3 for each practice for potentially offensive content as described in relation to RQ1) as dependent variables. The independent variables included in the models correspond to specific sub-RQs 2/3 and H1, H2. 

We chose ordinal logistic regression as the most appropriate model for the discrete ordinal dependent variables such as the Likert-scale survey responses as in the case of the present study. It has to be noted, however, that recent research suggests models such as GLM (generalized linear model) can be used with Likert-scale data as well \cite{harpe_how_2015}. The benefit of using GLM compared to ordinal logistic regression would be in the fact that it is easier to interpret. However, we opted for ordinal logistic regression as model diagnostics showed in our case several assumptions for GLM (specifically, linearity, homoskedasticity, and normality) were not met. Hence, the use of the GLM would have been inappropriate in this case. Ordinal logistic regression is not constrained by these assumptions. Instead, the assumptions for it include the absence of multicollinearity and proportional odds. We tested if the no multicollinearity assumption is met using VIF scores \cite{thompson_extracting_2017}. The goodness of fit of the models was assessed using an ordinal version of the Hosmer-Lemeshow test and the Lipsitz test \cite{fagerland_how_2017}. The proportional odds assumption for each model was first tested using Brant's test \cite{brant_assessing_1990}. However, this test is highly anticonservative - meaning that often the statistical significance of the test does not correspond to practical significance, especially when the number of predictors is high, sample size is large, or at least one continuous variable is used as a predictor \cite{allison_comparing_1999,peterson_partial_1990,kim_assessing_2003,das_application_2011}. Hence, in line with other research employing the methodology \cite{kim_assessing_2003,das_application_2011}, we have also used the graphical method to assess the practical significance of the assumption violation when Brant's test indicated statistical violation of the assumption. We discuss the models when this was the case and the implications for the interpretations of our findings at the end of this subsection.

\subsubsection{RQ2/3a: demographic characteristics}
In correspondence with RQ2/3a, we included the following independent variables on the demographic characteristics of the respondents: age (measured in numbers, data from the metadata on respondents collected and provided to us by Prolific), sex (binary category female/male\footnote{In addition to including a binary sex independent variable, we also included a non-binary gender variable (woman/man/non-binary). In the main text of the paper, we discuss only the models with sex as an independent variable - those allow us to contextualize our findings against those about CM attitudes on other platforms as those studies included sex, not gender, as an independent variable. However, we reran all our models using a non-binary gender variable instead of the binary sex variable. The models with gender are included in the Appendix, and the analysis shows that all our observations hold in those models as well.}, data collected and provided by Prolific), race (data collected and provided by Prolific; for the regression analysis we recoded the variable to a binary (White/non-White) variable), level of education (data collected and provided by Prolific).

\subsubsection{RQ2/3b: political leaning}
To measure the respondents' political leaning and include it as an independent variable in the model, we have used the following survey item adapted from \cite{kroh_measuring_2007}: "Political views are often seen as a spectrum between extremely liberal (left) to extremely conservative (right). Where would you place yourself on this scale where 0 means extremely liberal and 10 means extremely conservative?" 

\subsubsection{RQ2/3c, H1: trust in web search}
To measure the respondents' trust in web search and include it as an independent variable in the model to answer RQ2/3c and test H1, we used a composite measure of trust in search outputs adapted from \cite{stromback_news_2020}. The measure was constructed based on the survey items formulated as follows:

"Generally speaking, to what extent do you agree or disagree with the following statements about the information you find in web search engine results?
\begin{itemize}
    \item The selection of information I find in web search results tends to be fair and neutral
    \item The information I find in web search results tends to be accurate
    \item The information I find in web search results tends to be relevant for me"
\end{itemize}

The respondents could select their level of agreement with each of the statements on a 7-point Likert scale. Then, to construct the measure of the overall level of trust in web search, we calculated the mean of the responses to the three items (Cronbach's alpha = 0.789 indicating good item reliability).

\subsubsection{RQ2/3d: search engine use frequency}
To measure the frequency of use of specific search engines, we asked the respondents how often they use each of the search engines that are the most popular in the US \cite{statcounter_search_2022}: Google, Bing, Yahoo, Yandex, DuckDuckGo, Ecosia. The exact question was formulated as follows: "How often do you use each of the following search engines?" The responses were measured on a 7-point Likert scale.

\subsubsection{RQ2/3e, H2: search engines' independence}
To measure the degree to which the participants believe that search engines are independent of political or government influence, we have constructed a composite variable based on the mean of the participants' level of agreement (on a 7-point Likert scale) with each of the following two statements:
\begin{itemize}
    \item "Search engines are independent from undue political or government influence most of the time
    \item Search engines are independent from undue business or commercial influence most of the time"
\end{itemize}
This item was adapted from \cite{reuters_institute_resources_2016}. Cronbach's alpha = 
0.84 indicates good item reliability.

\subsubsection{A note on the violation of the proportional odds assumption}
As noted at the beginning of the subsection, we have relied on a combination of Brant's test \cite{brant_assessing_1990} and the graphical method to evaluate the practical and statistical significance of the violation of the proportional odds assumption across our models. There was a practically significant violation of the assumption for the Google use variable in the models where the dependent variable related to misleading content. Specifically, the graphical analysis indicated that more frequent use of Google is associated with \textit{slightly} lower likelihood of the users indicating that they somewhat support/support content moderation compared to strongly supporting it, and \textit{much} lower likelihood of them indicating that they oppose content moderation to any degree. This has to be taken into account when interpreting the findings. In addition, in an attempt to address this limitation, we have run partial proportional odds models \cite{peterson_partial_1990} that allow relaxing the proportional odds assumption for certain variables; however, these models indicated a very poor fit; hence, we opted not to use them. Instead, in addition to the models reported in the main text of the article, we have also run the models where content moderation preferences for misleading content are a dependent variable, omitting the Google use variable. These additional models are reported in the Appendix in Table \ref{table:coefficients_noGoogle}. Omitting Google use variable slightly changes the results - specifically, the Trust in SE variable has somewhat higher coefficients, indicating a stronger relationship to the DV, especially for the Reduction of reach of misleading information preferences, and in the case of this DV the use of DuckDuckGo emerges as a significant predictor when Google use is omitted. Since the results change only in minor ways, and the models with Google use omitted indicate a goodness-of-fit similar to those with Google use included, we opted for the inclusion of the full models with Google use included in the main text of the article to allow for consistent interpretation of the results. But here and below, in the results section, we emphasize the implications of the partial violation of proportional odds assumption for our findings.

\subsection{Ethics statement}
We obtained informed consent from the survey respondents for participation in the study and  informed the respondents about the goals of the study and the ways in which their data will be used. The full statement to which the respondents consented is available in the Appendix. The respondents were remunerated for participation in accordance with Prolific's terms (as the survey took around 15-20 minutes, we compensated the respondents with a 1/3 of the average hourly wage as determined by Prolific). We used only anonymized data and did not collect any personal information that would allow us or others to infer the identities of the respondents.

\section{Results}

\subsection{RQ1: User preferences for different content moderation options - descriptive analysis}

\begin{figure}[h]
  \centering
  \includegraphics[width=\linewidth]{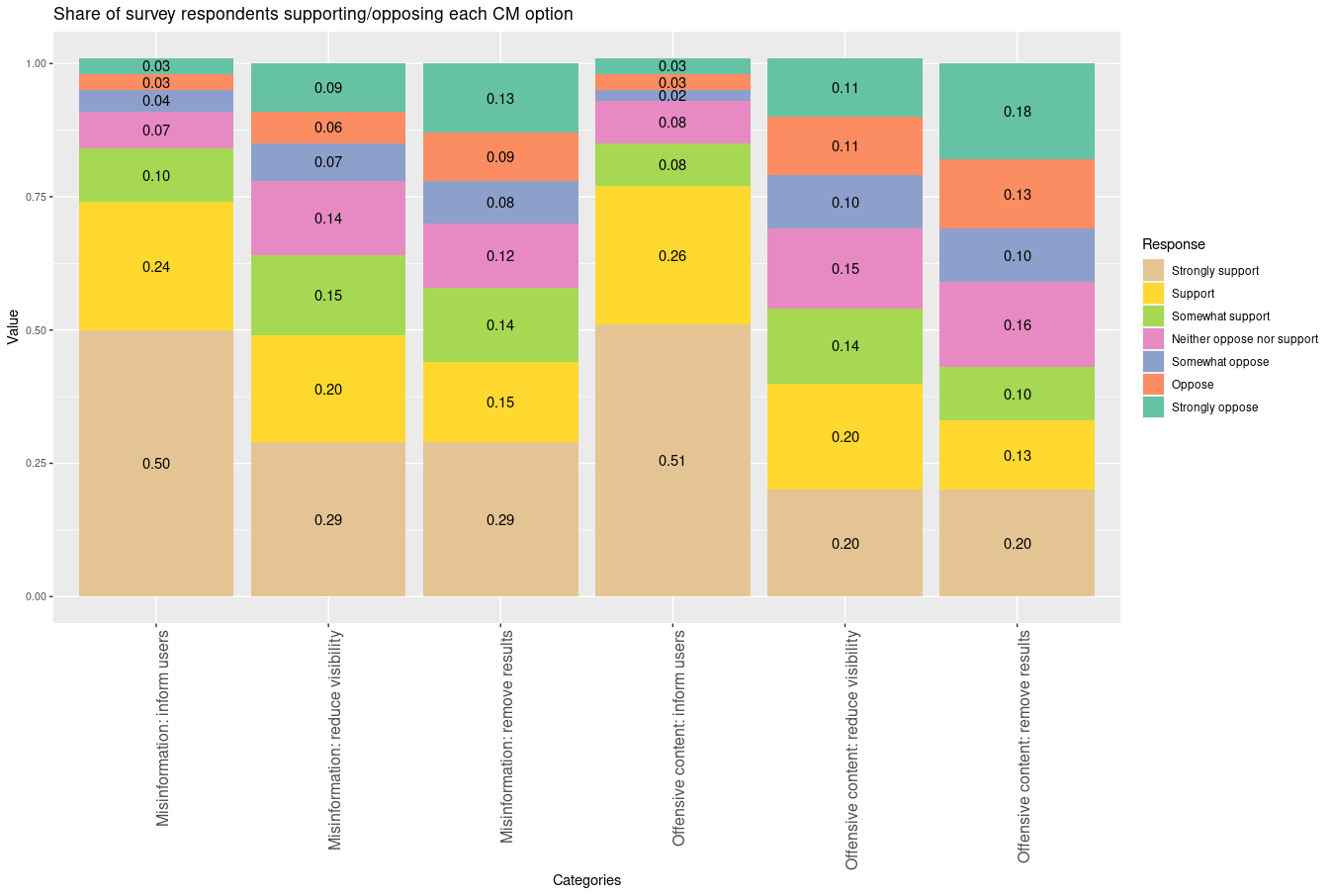}
  \caption{Share of survey respondents supporting/opposing each CM option.}
  \label{fig:modsupport}
\end{figure}

In Fig. \ref{fig:modsupport} we provide information on the shares of respondents supporting specific CM practices in web search. We observe that the option to inform users about potentially misleading or offensive content retrieved via web search is overwhelmingly supported with 84\% of respondents supporting\footnote{In this section we combine all support options - somewhat support/support/strongly support - to calculate overall support, same applies for the opposing options. A more fine-grained breakdown of the responses and corresponding share of survey participants selecting them is demonstrated in Fig. \ref{fig:modsupport}.} it for misleading content and 85\% for offensive content. Only 10\% of respondents oppose this option for misleading content and 8\% for offensive content.

Two other CM practices - to reduce the reach of certain content or to remove it from search results altogether - attracted less support from the respondents. For these practices, the shares of undecided users and those who only somewhat support/oppose the practice are higher than for informing users. Still, 64\% of respondents support reducing the reach of misleading content, and 54\% support reducing the reach of offensive content; the shares of respondents opposing this option is 22\% and 32\%, respectively. 58\% of survey participants also support removing misleading results from the outputs altogether, while 30\% oppose this option. In the case of offensive content, the removal of results seems to be a highly divisive issue - 43\% support this option, while 41\% oppose it. 

When it comes to the statistical significance of the observed discrepancies, the result of the Kruskall-Wallis test provided a p<0.00, indicating a statistically significant difference between user preferences for different types of CM. As shown in Table \ref{table:pairwise_wilcox}, the observed differences in the levels of support for different types of CM in web search when comparing different options pairwise are statistically significant for all pairs of options except informing users about misleading vs potentially offensive content and removing misleading results vs reducing the reach of potentially offensive content.

\begin{table}
\centering
\begin{tabular}{cccccc}
\toprule
& Mis: Inform & Mis: Reduce & Mis: Remove & Off: Inform & Off: Reduce \\
\midrule
Mis: Reduce & $< 0.00$ & - & - & - & - \\
Mis: Remove & $< 0.00$ & 0.00 & - & - & - \\
Off: Inform & 1.00 & $< 0.00$ & $< 0.00$ & - & - \\
Off: Reduce & $< 0.00$ & 0.00 & 1.00 & $< 0.00$ & - \\
Off: Remove & $< 0.00$ & 0.00 & 0.00 & $< 0.00$ & 0.00 \\
\bottomrule
\end{tabular}
\caption{P-values corresponding to pairwise comparisons of user preferences regarding different types of CM in web search (Wilcoxon signed rank test)}
\label{table:pairwise_wilcox}
\end{table}

We discuss the implications of our findings and how our observations correspond to the actual content moderation practices in web search in a dedicated Discussion section.

\subsection{RQs2,3, H1,2: Predictors of support for different CM practices}
In Table \ref{table:coefficients} we present the results of the regression analysis examining the relationship between the participants' characteristics and their support for different CM practices. The coefficients are exponentiated, and statistically significant coefficients are highlighted in red. The very bottom coefficients in Table \ref{table:coefficients} refer to the intercepts for each category of the dependent variable in the ordered logistic regression models. In ordinal regression, since the dependent variable has multiple ordered categories, separate intercepts are estimated for each category transition. These intercept coefficients provide information about the relative likelihood of being in each category compared to the reference category. They capture the inherent differences in the baseline odds of the different response categories before considering the effects of the predictor variables. For instance, the coefficients corresponding to 6|7 indicate the odds of the respondents selecting option 6 on the Likert scale ("Support") compared to option 7 ("Strongly support"). Since these coefficients are not relevant for our RQs and analysis, we do not interpret them below, and just keep them in the table for reference.

\begin{table}[!htbp]
\begin{center}
\begin{tabular}{l c c c c c c}
\hline
 & Mis: Inform & Mis: Reduce & Mis: Remove & Off: Inform & Off: Reduce & Off: Remove \\
\hline
Age              & $0.00$        & $-0.01$       & $0.01$        & $0.00$        & $-0.01$      & $0.00$        \\
                 & $(0.01)$      & $(0.01)$      & $(0.01)$      & $(0.01)$      &  $(0.01)$     & $(0.01)$      \\
Sex (Male)          & $0.08$        & $-0.22$       & $-0.15$       & $-0.02$       & $\textcolor{red}{-0.51^{**}}$ & $-0.24$       \\
                 & $(0.21)$      & $(0.20)$      & $(0.19)$      & $(0.21)$      & $(0.19)$     & $(0.19)$      \\
Education       & $-0.06$       & $-0.00$       & $-0.04$       & $0.01$        & $0.01$       & $0.11$        \\
                 & $(0.08)$      & $(0.07)$      & $(0.07)$      & $(0.08)$      & $(0.07)$     & $(0.07)$      \\
Ethnicity (White)  & $0.34$        & $-0.09$       & $-0.19$       & $0.28$        & $-0.20$      & $\textcolor{red}{-0.48^{*}}$   \\
                 & $(0.24)$      & $(0.22)$      & $(0.22)$      & $(0.24)$      & $(0.22)$     & $(0.22)$      \\
Trust in SE           & $\textcolor{red}{0.54^{***}}$  & $\textcolor{red}{0.28^{*}}$    & $0.15$        & $\textcolor{red}{0.24^{*}}$    & $0.04$       & $0.10$        \\
                 & $(0.13)$      & $(0.12)$      & $(0.12)$      & $(0.12)$      & $(0.11)$     & $(0.11)$      \\
SE independence            & $0.06$        & $\textcolor{red}{0.22^{**}}$   & $\textcolor{red}{0.26^{***}}$  & $0.14$        & $\textcolor{red}{0.31^{***}}$ & $\textcolor{red}{0.31^{***}}$  \\
                 & $(0.08)$      & $(0.07)$      & $(0.07)$      & $(0.08)$      & $(0.07)$     & $(0.07)$      \\
Political ideology\\ (left-right) & $\textcolor{red}{-0.25^{***}}$ & $\textcolor{red}{-0.21^{***}}$ & $\textcolor{red}{-0.21^{***}}$ & $\textcolor{red}{-0.15^{***}}$ & $\textcolor{red}{-0.10^{**}}$ & $-0.04$       \\
                 & $(0.04)$      & $(0.04)$      & $(0.04)$      & $(0.04)$      & $(0.04)$     & $(0.04)$      \\
Google use   & $\textcolor{red}{0.17^{*}}$    & $\textcolor{red}{0.28^{**}}$   & $\textcolor{red}{0.17^{*}}$    & $\textcolor{red}{0.20^{*}}$    & $0.13$       & $0.09$        \\
                 & $(0.08)$      & $(0.08)$      & $(0.08)$      & $(0.08)$      & $(0.08)$     & $(0.08)$      \\
DDG use     & $-0.06$       & $-0.09$       & $\textcolor{red}{-0.14^{*}}$   & $-0.08$       & $\textcolor{red}{-0.13^{*}}$  & $\textcolor{red}{-0.18^{***}}$ \\
                 & $(0.06)$      & $(0.05)$      & $(0.05)$      & $(0.06)$      & $(0.05)$     & $(0.05)$      \\
Yandex use   & $-0.13$       & $0.01$        & $-0.02$       & $-0.22$       & $-0.01$      & $0.02$        \\
                 & $(0.14)$      & $(0.14)$      & $(0.15)$      & $(0.14)$      & $(0.14)$     & $(0.14)$      \\
Yahoo  use   & $0.01$        & $0.06$        & $0.07$        & $0.03$        & $0.12$       & $\textcolor{red}{0.13^{*}}$    \\
                 & $(0.07)$      & $(0.07)$      & $(0.06)$      & $(0.07)$      & $(0.06)$     & $(0.06)$      \\
Bing use    & $0.10$        & $0.04$        & $0.05$        & $\textcolor{red}{0.12^{*}}$    & $0.04$       & $-0.03$       \\
                 & $(0.06)$      & $(0.06)$      & $(0.05)$      & $(0.06)$      & $(0.05)$     & $(0.05)$      \\
Ecosia use  & $-0.16$       & $-0.22$       & $-0.07$       & $-0.18$       & $-0.05$      & $0.12$        \\
                 & $(0.12)$      & $(0.13)$      & $(0.14)$      & $(0.12)$      & $(0.12)$     & $(0.13)$      \\
1|2              & $-0.86$       & $-0.43$       & $-0.51$       & $-1.57$       & $-1.46$      & $0.20$        \\
                 & $(0.91)$      & $(0.85)$      & $(0.84)$      & $(0.91)$      & $(0.83)$     & $(0.82)$      \\
2|3              & $-0.14$       & $0.33$        & $0.29$        & $-0.78$       & $-0.56$      & $1.00$        \\
                 & $(0.90)$      & $(0.85)$      & $(0.84)$      & $(0.89)$      & $(0.82)$     & $(0.82)$      \\
3|4              & $0.49$        & $0.93$        & $0.79$        & $-0.46$       & $0.04$       & $1.55$        \\
                 & $(0.90)$      & $(0.85)$      & $(0.84)$      & $(0.88)$      & $(0.82)$     & $(0.83)$      \\
4|5              & $1.24$        & $\textcolor{red}{1.78^{*}}$    & $1.37$        & $0.46$        & $0.78$       & $\textcolor{red}{2.30^{**}}$   \\
                 & $(0.90)$      & $(0.85)$      & $(0.84)$      & $(0.88)$      & $(0.82)$     & $(0.83)$      \\
5|6              & $\textcolor{red}{1.99^{*}}$    & $\textcolor{red}{2.52^{**}}$   & $\textcolor{red}{2.00^{*}}$    & $1.08$        & $1.42$       & $\textcolor{red}{2.78^{***}}$  \\
                 & $(0.90)$      & $(0.85)$      & $(0.84)$      & $(0.88)$      & $(0.82)$     & $(0.83)$      \\
6|7              & $\textcolor{red}{3.22^{***}}$  & $\textcolor{red}{3.44^{***}}$  & $\textcolor{red}{2.74^{**}}$   & $\textcolor{red}{2.37^{**}}$   & $\textcolor{red}{2.51^{**}}$  & $\textcolor{red}{3.48^{***}}$  \\
                 & $(0.90)$      & $(0.85)$      & $(0.85)$      & $(0.89)$      & $(0.83)$     & $(0.84)$      \\
\hline
AIC              & $1042.80$     & $1341.55$     & $1385.60$     & $1047.01$     & $1440.70$    & $1443.81$     \\
BIC              & $1118.11$     & $1416.86$     & $1460.91$     & $1122.32$     & $1516.00$    & $1519.12$     \\
Log Likelihood   & $-502.40$     & $-651.78$     & $-673.80$     & $-504.50$     & $-701.35$    & $-702.90$     \\
Deviance         & $1004.80$     & $1303.55$     & $1347.60$     & $1009.01$     & $1402.70$    & $1405.81$     \\
Num. obs.        & $389$         & $389$         & $389$         & $389$         & $389$        & $389$         \\
\hline
\multicolumn{7}{l}{\scriptsize{$^{***}p<0.001$; $^{**}p<0.01$; $^{*}p<0.05$}}
\end{tabular}
\caption{Outputs of regression models on the association between respondents' characteristics and their level of support for different CM practices for misleading (Mis) and Offensive (Off) content. Statistically significant coefficients are highlighted in red.}
\label{table:coefficients}
\end{center}
\end{table}

\subsubsection{RQs2/3a: support for CM and users' demographic characteristics}
We find no significant relationships between the age and level of education of the respondents and their levels of support for any CM practice in the context of web search. However, in a few cases, we observe a significant relationship between the respondents' sex and race and their support for CM for offensive content. Specifically, we find that male users are significantly more likely to oppose reducing the reach of potentially offensive content as compared to female users. Besides, White respondents are significantly more likely to oppose the removal of potentially offensive web search results than non-White ones.

\subsubsection{RQs2/3b: support for CM and users' political orientation}
We observe that respondents' political leaning is associated with their support for CM practices in all the examined cases with the exception of the removal of potentially offensive content. Similarly to the earlier observations in the context of social media \cite{kozyreva_free_2022,ballard_most_2019}, we find that more conservative users are less likely to support CM measures. This effect was stronger for misleading information than for offensive content.

\subsubsection{RQs2/3c, H1: support for CM and trust in web search results}
Based on earlier research about the relationship between trust in platforms and support for CM practices on these platforms, we hypothesized that the same relationship would be observed in the context of web search (H1). This hypothesis was only partially confirmed. Specifically, we find that trust is associated with increased support for informing users about both misleading and potentially offensive content with the effect being stronger for misleading content. Additionally, trust in web search results is related to the increased support for reducing the reach of misleading content; notably, this relationship is somewhat stronger in the models in which Google use is omitted (see Methodology and Table \ref{table:coefficients_noGoogle}). However, there was no association between trust in web search and support for removing results or reducing the reach of potentially offensive content.

\subsubsection{RQs2/3d: support for CM and usage of specific search engines}

\begin{figure}
  \centering
  \includegraphics[width=\linewidth]{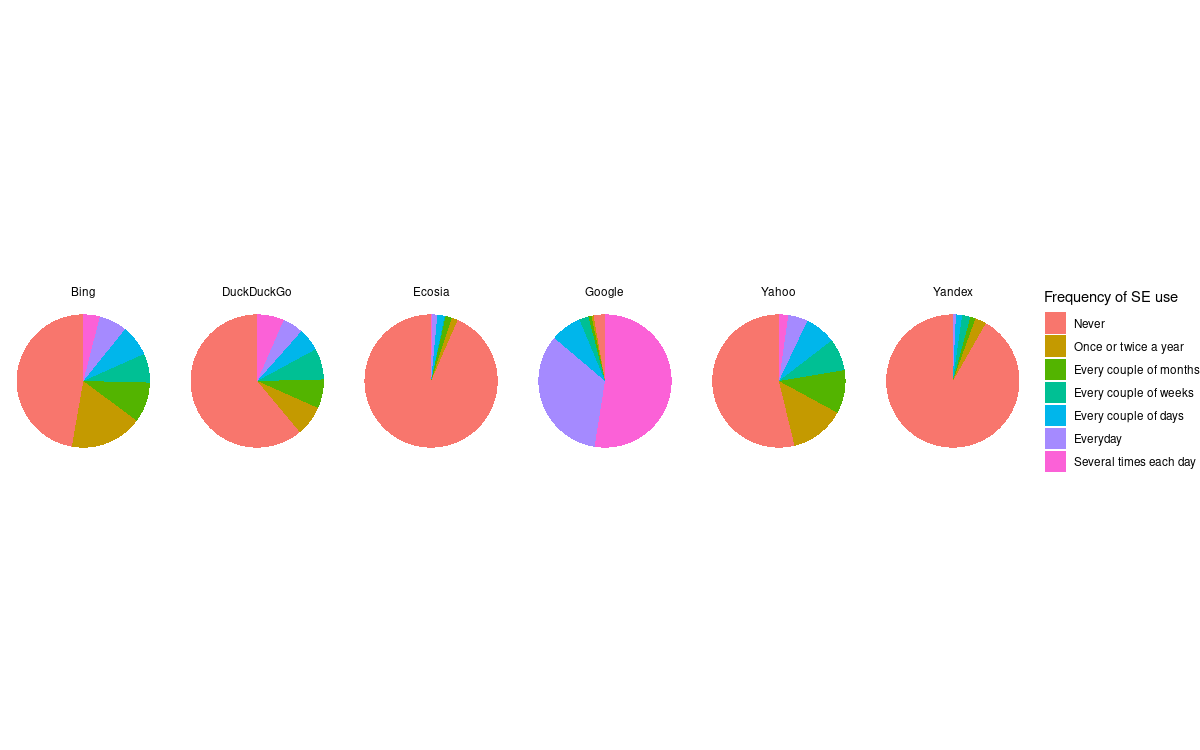}
  \caption{Distribution of the shares of respondents who report different frequencies of use of specific search engines.}
  \label{fig:searchuse}
\end{figure}

We observe multiple statistically significant associations between the use of specific web search engines and support for CM practices. It is worth noting, however, that the frequency of use of different SEs is drastically as one might expect based on the information about their respective market shares \cite{statcounter_search_2022}. We present the distribution of the frequencies of SEs' use in Fig.\ref{fig:searchuse}. Unsurprisingly, Google is the most used SE with almost all participants reporting using it at least a couple of times a year, and more than two-thirds stating they use it on a daily basis. Google is followed by Yahoo and Bing which are used at least once a year by around 50\% of the users, then comes DuckDuckGo with ca. 40\% respondents using it at least once a year. Ecosia and Yandex are used only by a small share of respondents. 

We find that Google use frequency is positively associated with support for most CM practices with the exception of the reduction of reach and removal of potentially offensive content. However, as noted in the methodology, it is necessary to interpret the coefficients with caution in this case due to the violation of the proportional odds assumption in the case of misleading content-related dependent variables. Specifically, our analysis during the model diagnostics stage revealed that more frequent use of Google is associated with \textit{slightly} lower likelihood of the users somewhat supporting/supporting content moderation compared to strongly supporting it, and \textit{much} lower likelihood of them opposing content moderation to any degree. Bing use frequency is positively related to the support for informing users about offensive content while Yahoo use frequency is associated with increased support for the removal of potentially offensive content. On the contrary, more frequent DuckDuckGo users are significantly less likely to support certain content moderation policies, specifically the removal of misleading or offensive content and the reduction of reach of the offensive content; when the use of Google variable is removed from the model, this relationship is also significant for the reduction of the reach of misleading content, see Table \ref{table:coefficients_noGoogle}. We find it important to highlight that all these observations emerge even when controlling for user demographics and political views, and discuss their implications in a dedicated section below.

\subsubsection{RQs2/3e, H2: support for CM and belief in the independence of SEs}
Our hypothesis (H2) that users' beliefs in the independence of search engines from political or business interests are associated with increased support for CM is confirmed in the case of the removal or reduction of reach of both misleading and offensive content. At the same time, there is no significant relationship between support for informing users about such content and belief in SE independence.

\section{Discussion}
Our observations show that there is a lot of divergence in the levels of the US adult respondents' support for different CM practices in web search for misleading or offensive content, with some of the differences explained by user characteristics, in particular political attitudes, frequency of SE use and trust in SE. 

\subsection{User attitudes to CM and actual SE moderation practices}
One CM practice that seems to be largely uncontroversial - as it is supported by an overwhelming majority of respondents - is informing users. Further, there is no statistically significant difference between the levels of user support for informing about misleading vs potentially offensive content. Currently, of the six most popular SEs, only Bing, Yandex and Google, according to their official statements and documents, inform users of some potentially problematic content (e.g., through dedicated warning labels), including misleading content. However, to the best of our knowledge, no such labels are attached to offensive content by any of the most popular SEs. This seems to be in clear contradiction with the US respondents' attitudes to CM: our analysis shows that informing users in the context of offensive content is supported even by slightly more respondents than informing users about misleading content (77\% vs 74\% of respondents) - albeit the difference is not statistically significant. Notably, respondents who use Bing and Google more frequently are more likely to support informing web search users about offensive content, suggesting that the implementation of such measures might be especially desired by the users of these two SEs. 

We observe another apparent contradiction between the level of support for CM practices in web search among our respondents and SE's actual practices. All aforementioned SEs except Ecosia remove search results altogether in certain cases - albeit mostly when it comes to explicitly illegal content - however, this practice is the one least supported by the respondents. While it still receives the support of a considerable share of respondents, the practice of removal is the only one to be opposed by over 10\% of users. Our analysis also reveals that the support for the complete removal of search results is significantly lower among more frequent users of DuckDuckGo for both misleading and offensive content. DuckDuckGo - at least based on the official statements and media reports we found - completely removes the results only when it comes to explicitly illegal content \cite{duckduckgo_news_nodate}. Thus, the engine's policies seem to be largely in alignment with the preferences of its more frequent users. The same applies in the case of downranking certain content: DuckDuckGo acknowledges the downgrading of "low quality" news websites \cite{duckduckgo_news_nodate} but does not mention downgrading offensive content, and explicitly states it does not downrank content based on its "truthfulness" \cite{duckduckgo_did_2023}. Its more frequent users are at the same time significantly more likely to oppose downranking offensive but not misleading content, which thus is to a degree in contradiction with the SE's practices.

\subsection{Predictors of user support for CM: web search vs social media}
Previous research has examined support for CM practices in the context of social media. We suggest it is worthwhile to compare our observations on CM attitudes in the context of web search to those regarding social media as such a comparison will reveal whether CM attitudes are similar across these different types of platforms.

We find that more conservative users are significantly less likely to support all forms of CM with the exception of the removal of offensive content than more liberal users. In the context of social media, scholars have observed a similar division in CM attitudes along ideological lines \cite{ballard_most_2019,kozyreva_free_2022} (though see \cite{riedl_antecedents_2022} that finds no association between partisanship and support for CM in the US). 

\cite{riedl_antecedents_2022} found no association between the US respondents' race or sex and preferences for content moderation on social media, while \cite{schoenebeck_youth_2021} showed that young women are more likely to support CM than young men. Our findings are broadly in line with both these observations. For most CM practices in web search, there is no association with the survey participants' sex and race. However, we find that men are significantly less likely than women to support the downgrading of offensive content in search outputs while White participants are significantly less likely than non-White respondents to support the complete removal of offensive results. We suggest this contextual difference might stem from the fact that women 
and non-White internet users 
encounter hate speech and other types of offensive content directed against them more often online \cite{CHETTY2018108}. Thus, they might be more in favor of reducing the reach or completely removing such content. Nonetheless, this explanation needs to be further examined and confirmed in future work. 

Our findings are in contrast to the observations of \cite{riedl_antecedents_2022} about the association between the users' age and education level and support for CM on social media. We find no such association in web search. It remains to be confirmed in future work that this is not due to the different operationalizations of CM (as the findings of \cite{riedl_antecedents_2022} contradict those of other scholars with regard to the relationship between social media CM attitudes and political ideology or sex).

Similarly to the findings of other scholars regarding support for CM on social media \cite{jhaver_did_2019,duffy_platform_2022,myers_west_censored_2018,saltz_encounters_2021}, we observe that higher trust in search engines and belief that they are independent from business or political influence are both significantly associated with stronger levels of support for CM practices. However, we find that only the reduction of reach of misleading content in search outputs is significantly related to both trust and belief in independence. Support for informing users about both misleading and offensive content is significantly related only to the general trust in search outputs while removing misleading and offensive content or reducing the reach of offensive content is significantly related only to the belief in the independence of SEs. One potential explanation is that even users who trust SEs support more drastic - in terms of the impact on information availability to search users - CM practices only if they also believe that SEs are independent and thus can make unbiased and fair decisions. This explanation would be in line with the observations regarding social media CM attitudes but is yet to be tested specifically in the context of web search.

To sum up, our findings are broadly in line with those regarding the support for CM on social media, suggesting that the mechanisms driving users' support and opposition to CM on online platforms are similar across different platform types.

\subsection{Limitations and future work}
Our study is not without limitations. First, we looked only at the respondents from the US thus our findings hold only for this context. While this allowed us to contextualize our findings against the studies about social media CM attitudes that were mainly conducted in the US, we believe it is necessary to examine attitudes to CM in other contexts as well, preferably through comparative analysis in order to draw more general and meaningful conclusions. We suggest that a comparative analysis of CM attitudes on both social media and SEs - and possibly other types of platforms - across national contexts would be a particularly fruitful direction for future work. Second, we did not present the survey respondents with specific definitions or examples of either misleading or offensive content. We did it on purpose to gauge the respondents' general attitudes to CM, relying on their own perceptions of what constitutes misleading or offensive information. However, research demonstrates that users can have different views on whether certain content is misleading or offensive \cite{RUOKOLAINEN2020102127}. Our study does not account for such differences but we suggest that it would be important to examine how they are related to support for CM practices in the future. The latter limitation is especially relevant in the context of the actual implementation of CM by search engines. Even if SEs, for instance, start informing users about misleading or offensive content - as there is broad support for this measure as we show - defining what constitutes such content and harmonizing this definition taking into account potentially diverging opinions of different groups of users will be a major challenge. We suggest it would also be important in future work not only to examine what different users perceive as offensive or misleading but also to examine different mechanisms that would allow for the implementation of CM in a way that is both supported by diverse groups of users and is conducive to fostering well-informed society. In addition, since users' declared preferences might not always match their actual behavior, we suggest it would be worthwhile in future work to evaluate how users in fact perceive more or less moderated search results. This can be done, for example, by relying on experimental methods. Finally, in this paper, we have only explored and described users' preferences towards content moderation in web search and examined their predictors. Future work could additionally explore what are the most effective moderation measures in web search and whether or not users' preferences are in alignment with the most effective techniques. We highlight that while user preferences should be taken into account when designing moderation policies, they do not necessarily always correspond to what the actual most effective practices for moderation would be.  Thus, for content moderation design it would be inappropriate to simply reflect user preferences - rather, it would be worthwhile to further explore them and their consequences and engage with them critically.

\section{Conclusion}
In this paper, we aimed to address the gap in the understanding of public attitudes to CM practices in web search in application to potentially misleading and potentially offensive content based on a survey of a representative sample of the US adult population. In addition to examining the user attitudes to different content moderation practices, we have first conducted an overview of the actual practices employed by different search engines and systematized them, identifying three main practices: informing the users about certain types of content; reducing the reach of certain content; removing certain content altogether. In terms of user attitudes towards these practices, we find that there is broad support for informing users about misleading/offensive content among the respondents. The attitudes towards reducing the reach of such information through downgrading it in search results and completely removing such information are more divided. While high shares of respondents are in support of these practices, the support is not as broad as for informing users. Further, over 10\% of respondents strongly oppose removing search results altogether. We also find that levels of support for content moderation are significantly associated with the respondents' political ideology - more conservative users are less likely to support CM practices - and trust in web search as well as belief in the independence of SEs - users who trust SEs more and have a stronger belief in their independence are more likely to support CM in search. In addition, we find that male users are less likely to support the downgrading of potentially offensive information in search results, while White users are less likely to support its complete removal. Our findings on the associations between user characteristics and attitudes to content moderation in web search are broadly in line with those previously made by scholars in the context of social media. 

\bibliographystyle{ACM-Reference-Format}

\begin{thebibliography}{89}


\ifx \showCODEN    \undefined \def \showCODEN     #1{\unskip}     \fi
\ifx \showDOI      \undefined \def \showDOI       #1{#1}\fi
\ifx \showISBNx    \undefined \def \showISBNx     #1{\unskip}     \fi
\ifx \showISBNxiii \undefined \def \showISBNxiii  #1{\unskip}     \fi
\ifx \showISSN     \undefined \def \showISSN      #1{\unskip}     \fi
\ifx \showLCCN     \undefined \def \showLCCN      #1{\unskip}     \fi
\ifx \shownote     \undefined \def \shownote      #1{#1}          \fi
\ifx \showarticletitle \undefined \def \showarticletitle #1{#1}   \fi
\ifx \showURL      \undefined \def \showURL       {\relax}        \fi
\providecommand\bibfield[2]{#2}
\providecommand\bibinfo[2]{#2}
\providecommand\natexlab[1]{#1}
\providecommand\showeprint[2][]{arXiv:#2}

\bibitem[Alizadeh et~al\mbox{.}(2022)]%
        {alizadeh_content_2022}
\bibfield{author}{\bibinfo{person}{Meysam Alizadeh}, \bibinfo{person}{Fabrizio Gilardi}, \bibinfo{person}{Emma Hoes}, \bibinfo{person}{K.~Jonathan Klüser}, \bibinfo{person}{Maël Kubli}, {and} \bibinfo{person}{Nahema Marchal}.} \bibinfo{year}{2022}\natexlab{}.
\newblock \showarticletitle{Content {Moderation} {As} a {Political} {Issue}: {The} {Twitter} {Discourse} {Around} {Trump}'s {Ban}}.
\newblock \bibinfo{journal}{\emph{Journal of Quantitative Description: Digital Media}}  \bibinfo{volume}{2} (\bibinfo{date}{Oct.} \bibinfo{year}{2022}).
\newblock
\showISSN{2673-8813}
\urldef\tempurl%
\url{https://doi.org/10.51685/jqd.2022.023}
\showDOI{\tempurl}


\bibitem[Allison(1999)]%
        {allison_comparing_1999}
\bibfield{author}{\bibinfo{person}{Paul~D. Allison}.} \bibinfo{year}{1999}\natexlab{}.
\newblock \showarticletitle{Comparing {Logit} and {Probit} {Coefficients} {Across} {Groups}}.
\newblock \bibinfo{journal}{\emph{Sociological Methods \& Research}} \bibinfo{volume}{28}, \bibinfo{number}{2} (\bibinfo{date}{Nov.} \bibinfo{year}{1999}), \bibinfo{pages}{186--208}.
\newblock
\showISSN{0049-1241}
\urldef\tempurl%
\url{https://doi.org/10.1177/0049124199028002003}
\showDOI{\tempurl}
\newblock
\shownote{Publisher: SAGE Publications Inc}.


\bibitem[Arendt et~al\mbox{.}(2020)]%
        {arendt_investigating_2020}
\bibfield{author}{\bibinfo{person}{Florian Arendt}, \bibinfo{person}{Mario Haim}, {and} \bibinfo{person}{Sebastian Scherr}.} \bibinfo{year}{2020}\natexlab{}.
\newblock \showarticletitle{Investigating {Google}'s suicide-prevention efforts in celebrity suicides using agent-based testing: {A} cross-national study in four {European} countries}.
\newblock \bibinfo{journal}{\emph{Social Science \& Medicine}}  \bibinfo{volume}{262} (\bibinfo{date}{Oct.} \bibinfo{year}{2020}), \bibinfo{pages}{112692}.
\newblock
\showISSN{0277-9536}
\urldef\tempurl%
\url{https://doi.org/10.1016/j.socscimed.2019.112692}
\showDOI{\tempurl}


\bibitem[Atreja et~al\mbox{.}(2022)]%
        {atreja_what_2022}
\bibfield{author}{\bibinfo{person}{Shubham Atreja}, \bibinfo{person}{Libby Hemphill}, {and} \bibinfo{person}{Paul Resnick}.} \bibinfo{year}{2022}\natexlab{}.
\newblock \bibinfo{title}{What is the {Will} of the {People}? {Moderation} {Preferences} for {Misinformation}}.
\newblock
\newblock
\urldef\tempurl%
\url{https://doi.org/10.48550/arXiv.2202.00799}
\showDOI{\tempurl}
\newblock
\shownote{arXiv:2202.00799 [cs]}.


\bibitem[Ballard(2019)]%
        {ballard_most_2019}
\bibfield{author}{\bibinfo{person}{Jamie Ballard}.} \bibinfo{year}{2019}\natexlab{}.
\newblock \bibinfo{title}{Most conservatives believe removing content and comments on social media is suppressing free speech {\textbar} {YouGov}}.
\newblock
\newblock
\urldef\tempurl%
\url{https://today.yougov.com/topics/technology/articles-reports/2019/04/29/content-moderation-social-media-free-speech-poll}
\showURL{%
\tempurl}


\bibitem[Bilić(2016)]%
        {bilic_search_2016}
\bibfield{author}{\bibinfo{person}{Paško Bilić}.} \bibinfo{year}{2016}\natexlab{}.
\newblock \showarticletitle{Search algorithms, hidden labour and information control}.
\newblock \bibinfo{journal}{\emph{Big Data \& Society}} \bibinfo{volume}{3}, \bibinfo{number}{1} (\bibinfo{date}{June} \bibinfo{year}{2016}), \bibinfo{pages}{2053951716652159}.
\newblock
\showISSN{2053-9517}
\urldef\tempurl%
\url{https://doi.org/10.1177/2053951716652159}
\showDOI{\tempurl}
\newblock
\shownote{Publisher: SAGE Publications Ltd}.


\bibitem[Brant(1990)]%
        {brant_assessing_1990}
\bibfield{author}{\bibinfo{person}{Rollin Brant}.} \bibinfo{year}{1990}\natexlab{}.
\newblock \showarticletitle{Assessing {Proportionality} in the {Proportional} {Odds} {Model} for {Ordinal} {Logistic} {Regression}}.
\newblock \bibinfo{journal}{\emph{Biometrics}} \bibinfo{volume}{46}, \bibinfo{number}{4} (\bibinfo{date}{Dec.} \bibinfo{year}{1990}), \bibinfo{pages}{1171}.
\newblock
\showISSN{0006341X}
\urldef\tempurl%
\url{https://doi.org/10.2307/2532457}
\showDOI{\tempurl}


\bibitem[Calleberg(2021)]%
        {calleberg_making_2021}
\bibfield{author}{\bibinfo{person}{Erik Calleberg}.} \bibinfo{year}{2021}\natexlab{}.
\newblock \bibinfo{booktitle}{\emph{Making {Content} {Moderation} {Less} {Frustrating} : {How} {Do} {Users} {Experience} {Explanatory} {Human} and {AI} {Moderation} {Messages}}}.
\newblock
\urldef\tempurl%
\url{https://urn.kb.se/resolve?urn=urn:nbn:se:sh:diva-46050}
\showURL{%
\tempurl}


\bibitem[Chetty and Alathur(2018)]%
        {CHETTY2018108}
\bibfield{author}{\bibinfo{person}{Naganna Chetty} {and} \bibinfo{person}{Sreejith Alathur}.} \bibinfo{year}{2018}\natexlab{}.
\newblock \showarticletitle{Hate speech review in the context of online social networks}.
\newblock \bibinfo{journal}{\emph{Aggression and Violent Behavior}}  \bibinfo{volume}{40} (\bibinfo{year}{2018}), \bibinfo{pages}{108--118}.
\newblock
\showISSN{1359-1789}
\urldef\tempurl%
\url{https://doi.org/10.1016/j.avb.2018.05.003}
\showDOI{\tempurl}


\bibitem[core team(2023)]%
        {r_core_team_stats-package_2023}
\bibfield{author}{\bibinfo{person}{R core team}.} \bibinfo{year}{2023}\natexlab{}.
\newblock \bibinfo{title}{stats-package: {The} {R} {Stats} {Package}}.
\newblock
\newblock
\urldef\tempurl%
\url{https://rdrr.io/r/stats/stats-package.html}
\showURL{%
\tempurl}


\bibitem[Cuan-Baltazar et~al\mbox{.}(2020)]%
        {info:doi/10.2196/18444}
\bibfield{author}{\bibinfo{person}{Jose~Yunam Cuan-Baltazar}, \bibinfo{person}{Maria~Jos{\'e} Mu{\~{n}}oz-Perez}, \bibinfo{person}{Carolina Robledo-Vega}, \bibinfo{person}{Maria~Fernanda P{\'e}rez-Zepeda}, {and} \bibinfo{person}{Elena Soto-Vega}.} \bibinfo{year}{2020}\natexlab{}.
\newblock \showarticletitle{Misinformation of COVID-19 on the Internet: Infodemiology Study}.
\newblock \bibinfo{journal}{\emph{JMIR Public Health Surveill}} \bibinfo{volume}{6}, \bibinfo{number}{2} (\bibinfo{date}{9 Apr} \bibinfo{year}{2020}), \bibinfo{pages}{e18444}.
\newblock
\showISSN{2369-2960}
\urldef\tempurl%
\url{https://doi.org/10.2196/18444}
\showDOI{\tempurl}


\bibitem[Das and Rahman(2011)]%
        {das_application_2011}
\bibfield{author}{\bibinfo{person}{Sumonkanti Das} {and} \bibinfo{person}{Rajwanur~M. Rahman}.} \bibinfo{year}{2011}\natexlab{}.
\newblock \showarticletitle{Application of ordinal logistic regression analysis in determining risk factors of child malnutrition in {Bangladesh}}.
\newblock \bibinfo{journal}{\emph{Nutrition Journal}} \bibinfo{volume}{10}, \bibinfo{number}{1} (\bibinfo{date}{Nov.} \bibinfo{year}{2011}), \bibinfo{pages}{124}.
\newblock
\showISSN{1475-2891}
\urldef\tempurl%
\url{https://doi.org/10.1186/1475-2891-10-124}
\showDOI{\tempurl}


\bibitem[DuckDuckGo(2022)]%
        {duckduckgo_news_nodate}
\bibfield{author}{\bibinfo{person}{DuckDuckGo}.} \bibinfo{year}{2022}\natexlab{}.
\newblock \bibinfo{title}{News {Rankings}}.
\newblock
\newblock
\urldef\tempurl%
\url{https://help.duckduckgo.com/duckduckgo-help-pages/results/news-rankings/}
\showURL{%
\tempurl}


\bibitem[DuckDuckGo(2023a)]%
        {duckduckgo_did_2023}
\bibfield{author}{\bibinfo{person}{DuckDuckGo}.} \bibinfo{year}{2023}\natexlab{a}.
\newblock \bibinfo{title}{Did {DuckDuckGo} censor search results about the {Russia}-{Ukraine} war?}
\newblock
\newblock
\urldef\tempurl%
\url{https://duckduckgo.com/duckduckgo-help-pages/misconceptions/did-duckduckgo-censor-russian-ukraine-war-search-results/}
\showURL{%
\tempurl}


\bibitem[DuckDuckGo(2023b)]%
        {duckduckgo_does_nodate}
\bibfield{author}{\bibinfo{person}{DuckDuckGo}.} \bibinfo{year}{2023}\natexlab{b}.
\newblock \bibinfo{title}{Does {DuckDuckGo} censor or otherwise politically bias their search results?}
\newblock
\newblock
\urldef\tempurl%
\url{https://duckduckgo.com/duckduckgo-help-pages/misconceptions/does-duckduckgo-censor-search-results/}
\showURL{%
\tempurl}


\bibitem[Duffy and Meisner(2022)]%
        {duffy_platform_2022}
\bibfield{author}{\bibinfo{person}{Brooke~Erin Duffy} {and} \bibinfo{person}{Colten Meisner}.} \bibinfo{year}{2022}\natexlab{}.
\newblock \showarticletitle{Platform governance at the margins: {Social} media creators’ experiences with algorithmic (in)visibility}.
\newblock \bibinfo{journal}{\emph{Media, Culture \& Society}} (\bibinfo{date}{July} \bibinfo{year}{2022}), \bibinfo{pages}{01634437221111923}.
\newblock
\showISSN{0163-4437}
\urldef\tempurl%
\url{https://doi.org/10.1177/01634437221111923}
\showDOI{\tempurl}
\newblock
\shownote{Publisher: SAGE Publications Ltd}.


\bibitem[Edelman(2021)]%
        {edelman_2021_2021}
\bibfield{author}{\bibinfo{person}{Edelman}.} \bibinfo{year}{2021}\natexlab{}.
\newblock \bibinfo{title}{2021 {Edelman} {Trust} {Barometer}}.
\newblock
\newblock
\urldef\tempurl%
\url{https://www.edelman.com/trust/2021-trust-barometer}
\showURL{%
\tempurl}


\bibitem[Epstein and Robertson(2015)]%
        {epstein_search_2015}
\bibfield{author}{\bibinfo{person}{Robert Epstein} {and} \bibinfo{person}{Ronald~E. Robertson}.} \bibinfo{year}{2015}\natexlab{}.
\newblock \showarticletitle{The search engine manipulation effect ({SEME}) and its possible impact on the outcomes of elections}.
\newblock \bibinfo{journal}{\emph{Proceedings of the National Academy of Sciences}} \bibinfo{volume}{112}, \bibinfo{number}{33} (\bibinfo{date}{Aug.} \bibinfo{year}{2015}), \bibinfo{pages}{E4512--E4521}.
\newblock
\urldef\tempurl%
\url{https://doi.org/10.1073/pnas.1419828112}
\showDOI{\tempurl}
\newblock
\shownote{Publisher: Proceedings of the National Academy of Sciences}.


\bibitem[Fagerland and Hosmer(2017)]%
        {fagerland_how_2017}
\bibfield{author}{\bibinfo{person}{Morten~W. Fagerland} {and} \bibinfo{person}{David~W. Hosmer}.} \bibinfo{year}{2017}\natexlab{}.
\newblock \showarticletitle{How to {Test} for {Goodness} of {Fit} in {Ordinal} {Logistic} {Regression} {Models}}.
\newblock \bibinfo{journal}{\emph{The Stata Journal: Promoting communications on statistics and Stata}} \bibinfo{volume}{17}, \bibinfo{number}{3} (\bibinfo{date}{Sept.} \bibinfo{year}{2017}), \bibinfo{pages}{668--686}.
\newblock
\showISSN{1536-867X, 1536-8734}
\urldef\tempurl%
\url{https://doi.org/10.1177/1536867X1701700308}
\showDOI{\tempurl}


\bibitem[Fisher et~al\mbox{.}(2015)]%
        {fisher_searching_2015}
\bibfield{author}{\bibinfo{person}{Matthew Fisher}, \bibinfo{person}{Mariel~K. Goddu}, {and} \bibinfo{person}{Frank~C. Keil}.} \bibinfo{year}{2015}\natexlab{}.
\newblock \showarticletitle{Searching for explanations: {How} the {Internet} inflates estimates of internal knowledge.}
\newblock \bibinfo{journal}{\emph{Journal of Experimental Psychology: General}} \bibinfo{volume}{144}, \bibinfo{number}{3} (\bibinfo{date}{June} \bibinfo{year}{2015}), \bibinfo{pages}{674--687}.
\newblock
\showISSN{1939-2222, 0096-3445}
\urldef\tempurl%
\url{https://doi.org/10.1037/xge0000070}
\showDOI{\tempurl}


\bibitem[French et~al\mbox{.}(2008)]%
        {french_multivariate_nodate}
\bibfield{author}{\bibinfo{person}{Aaron French}, \bibinfo{person}{Marcelo Macedo}, \bibinfo{person}{John Poulsen}, \bibinfo{person}{Tyler Waterson}, {and} \bibinfo{person}{Angela Yu}.} \bibinfo{year}{2008}\natexlab{}.
\newblock \showarticletitle{Multivariate {Analysis} of {Variance} ({MANOVA})}.
\newblock  (\bibinfo{year}{2008}).
\newblock


\bibitem[{Gabriel Weinberg [@yegg]}(2022)]%
        {gabriel_weinberg_yegg_like_2022}
\bibfield{author}{\bibinfo{person}{{Gabriel Weinberg [@yegg]}}.} \bibinfo{year}{2022}\natexlab{}.
\newblock \bibinfo{title}{Like so many others I am sickened by Russia's invasion of Ukraine and the gigantic humanitarian crisis it continues to create. StandWithUkraine At DuckDuckGo we've been rolling out search updates that down-rank sitess associated with Russian disinformation}.
\newblock
\newblock
\urldef\tempurl%
\url{https://twitter.com/yegg/status/1501716484761997318}
\showURL{%
\tempurl}


\bibitem[Ganesh and Bright(2020)]%
        {ganesh_countering_2020}
\bibfield{author}{\bibinfo{person}{Bharath Ganesh} {and} \bibinfo{person}{Jonathan Bright}.} \bibinfo{year}{2020}\natexlab{}.
\newblock \showarticletitle{Countering {Extremists} on {Social} {Media}: {Challenges} for {Strategic} {Communication} and {Content} {Moderation}}.
\newblock \bibinfo{journal}{\emph{Policy \& Internet}} \bibinfo{volume}{12}, \bibinfo{number}{1} (\bibinfo{year}{2020}), \bibinfo{pages}{6--19}.
\newblock
\showISSN{1944-2866}
\urldef\tempurl%
\url{https://doi.org/10.1002/poi3.236}
\showDOI{\tempurl}
\newblock
\shownote{\_eprint: https://onlinelibrary.wiley.com/doi/pdf/10.1002/poi3.236}.


\bibitem[Gerrard(2018)]%
        {gerrard_beyond_2018}
\bibfield{author}{\bibinfo{person}{Ysabel Gerrard}.} \bibinfo{year}{2018}\natexlab{}.
\newblock \showarticletitle{Beyond the hashtag: {Circumventing} content moderation on social media}.
\newblock \bibinfo{journal}{\emph{New Media \& Society}} \bibinfo{volume}{20}, \bibinfo{number}{12} (\bibinfo{date}{Dec.} \bibinfo{year}{2018}), \bibinfo{pages}{4492--4511}.
\newblock
\showISSN{1461-4448}
\urldef\tempurl%
\url{https://doi.org/10.1177/1461444818776611}
\showDOI{\tempurl}
\newblock
\shownote{Publisher: SAGE Publications}.


\bibitem[Ghenai(2017)]%
        {10.1145/3079452.3079483}
\bibfield{author}{\bibinfo{person}{Amira Ghenai}.} \bibinfo{year}{2017}\natexlab{}.
\newblock \showarticletitle{Health Misinformation in Search and Social Media}. In \bibinfo{booktitle}{\emph{Proceedings of the 2017 International Conference on Digital Health}} (London, United Kingdom) \emph{(\bibinfo{series}{DH '17})}. \bibinfo{publisher}{Association for Computing Machinery}, \bibinfo{address}{New York, NY, USA}, \bibinfo{pages}{235–236}.
\newblock
\showISBNx{9781450352499}
\urldef\tempurl%
\url{https://doi.org/10.1145/3079452.3079483}
\showDOI{\tempurl}


\bibitem[Gillespie(2018)]%
        {gillespie_custodians_2018}
\bibfield{author}{\bibinfo{person}{Tarleton Gillespie}.} \bibinfo{year}{2018}\natexlab{}.
\newblock \bibinfo{booktitle}{\emph{Custodians of the {Internet}: {Platforms}, {Content} {Moderation}, and the {Hidden} {Decisions} {That} {Shape} {Social} {Media}} (\bibinfo{edition}{illustrated edition} ed.)}.
\newblock \bibinfo{publisher}{Yale University Press}, \bibinfo{address}{New Haven}.
\newblock
\showISBNx{978-0-300-17313-0}


\bibitem[Gillespie(2020)]%
        {gillespie_content_2020}
\bibfield{author}{\bibinfo{person}{Tarleton Gillespie}.} \bibinfo{year}{2020}\natexlab{}.
\newblock \showarticletitle{Content moderation, {AI}, and the question of scale}.
\newblock \bibinfo{journal}{\emph{Big Data \& Society}} \bibinfo{volume}{7}, \bibinfo{number}{2} (\bibinfo{date}{July} \bibinfo{year}{2020}), \bibinfo{pages}{2053951720943234}.
\newblock
\showISSN{2053-9517}
\urldef\tempurl%
\url{https://doi.org/10.1177/2053951720943234}
\showDOI{\tempurl}
\newblock
\shownote{Publisher: SAGE Publications Ltd}.


\bibitem[Gillespie(2022)]%
        {gillespie_not_2022}
\bibfield{author}{\bibinfo{person}{Tarleton Gillespie}.} \bibinfo{year}{2022}\natexlab{}.
\newblock \showarticletitle{Do {Not} {Recommend}? {Reduction} as a {Form} of {Content} {Moderation}}.
\newblock \bibinfo{journal}{\emph{Social Media + Society}} \bibinfo{volume}{8}, \bibinfo{number}{3} (\bibinfo{date}{July} \bibinfo{year}{2022}), \bibinfo{pages}{20563051221117552}.
\newblock
\showISSN{2056-3051}
\urldef\tempurl%
\url{https://doi.org/10.1177/20563051221117552}
\showDOI{\tempurl}
\newblock
\shownote{Publisher: SAGE Publications Ltd}.


\bibitem[Google(2020)]%
        {google_information_2020}
\bibfield{author}{\bibinfo{person}{Google}.} \bibinfo{year}{2020}\natexlab{}.
\newblock \bibinfo{booktitle}{\emph{Information {Quality} \& {Content} {Moderation}}}.
\newblock \bibinfo{type}{{T}echnical {R}eport}.
\newblock
\urldef\tempurl%
\url{https://blog.google/documents/83/information_quality_content_moderation_white_paper.pdf/}
\showURL{%
\tempurl}


\bibitem[Google(2022a)]%
        {google_general_2022}
\bibfield{author}{\bibinfo{person}{Google}.} \bibinfo{year}{2022}\natexlab{a}.
\newblock \bibinfo{booktitle}{\emph{General {Search} {Quality} {Rating} {Guidelines}}}.
\newblock \bibinfo{type}{{T}echnical {R}eport}.
\newblock
\urldef\tempurl%
\url{https://static.googleusercontent.com/media/guidelines.raterhub.com/en//searchqualityevaluatorguidelines.pdf}
\showURL{%
\tempurl}


\bibitem[Google(2022b)]%
        {google_manage_2022}
\bibfield{author}{\bibinfo{person}{Google}.} \bibinfo{year}{2022}\natexlab{b}.
\newblock \bibinfo{title}{Manage warnings about unsafe sites - {Android} - {Google} {Chrome} {Help}}.
\newblock
\newblock
\urldef\tempurl%
\url{https://support.google.com/chrome/answer/99020?hl=en}
\showURL{%
\tempurl}


\bibitem[Google(2022c)]%
        {google_rigorous_2022}
\bibfield{author}{\bibinfo{person}{Google}.} \bibinfo{year}{2022}\natexlab{c}.
\newblock \bibinfo{title}{Rigorous {Testing} – {How} {Google} {Search} {Works}}.
\newblock
\newblock
\urldef\tempurl%
\url{https://www.google.com/search/howsearchworks/how-search-works/rigorous-testing/}
\showURL{%
\tempurl}


\bibitem[Google(2022d)]%
        {google_search_2022}
\bibfield{author}{\bibinfo{person}{Google}.} \bibinfo{year}{2022}\natexlab{d}.
\newblock \bibinfo{booktitle}{\emph{Search {Quality} {Rater} {Guidelines}: {An} {Overview}}}.
\newblock \bibinfo{type}{{T}echnical {R}eport}.
\newblock
\urldef\tempurl%
\url{https://services.google.com/fh/files/misc/hsw-sqrg.pdf}
\showURL{%
\tempurl}


\bibitem[Gorwa et~al\mbox{.}(2020)]%
        {gorwa_algorithmic_2020}
\bibfield{author}{\bibinfo{person}{Robert Gorwa}, \bibinfo{person}{Reuben Binns}, {and} \bibinfo{person}{Christian Katzenbach}.} \bibinfo{year}{2020}\natexlab{}.
\newblock \showarticletitle{Algorithmic content moderation: {Technical} and political challenges in the automation of platform governance}.
\newblock \bibinfo{journal}{\emph{Big Data \& Society}} \bibinfo{volume}{7}, \bibinfo{number}{1} (\bibinfo{date}{Jan.} \bibinfo{year}{2020}), \bibinfo{pages}{2053951719897945}.
\newblock
\showISSN{2053-9517}
\urldef\tempurl%
\url{https://doi.org/10.1177/2053951719897945}
\showDOI{\tempurl}
\newblock
\shownote{Publisher: SAGE Publications Ltd}.


\bibitem[Hargittai et~al\mbox{.}(2010)]%
        {hargittai_trust_2010}
\bibfield{author}{\bibinfo{person}{Eszter Hargittai}, \bibinfo{person}{Lindsay Fullerton}, \bibinfo{person}{Ericka Menchen-Trevino}, {and} \bibinfo{person}{Kristin~Yates Thomas}.} \bibinfo{year}{2010}\natexlab{}.
\newblock \showarticletitle{Trust {Online}: {Young} {Adults}' {Evaluation} of {Web} {Content}}.
\newblock \bibinfo{journal}{\emph{International Journal of Communication}} \bibinfo{volume}{4}, \bibinfo{number}{0} (\bibinfo{date}{April} \bibinfo{year}{2010}), \bibinfo{pages}{27}.
\newblock
\showISSN{1932-8036}
\urldef\tempurl%
\url{https://ijoc.org/index.php/ijoc/article/view/636}
\showURL{%
\tempurl}
\newblock
\shownote{Number: 0}.


\bibitem[Harpe(2015)]%
        {harpe_how_2015}
\bibfield{author}{\bibinfo{person}{Spencer~E. Harpe}.} \bibinfo{year}{2015}\natexlab{}.
\newblock \showarticletitle{How to analyze {Likert} and other rating scale data}.
\newblock \bibinfo{journal}{\emph{Currents in Pharmacy Teaching and Learning}} \bibinfo{volume}{7}, \bibinfo{number}{6} (\bibinfo{date}{Nov.} \bibinfo{year}{2015}), \bibinfo{pages}{836--850}.
\newblock
\showISSN{1877-1297}
\urldef\tempurl%
\url{https://doi.org/10.1016/j.cptl.2015.08.001}
\showDOI{\tempurl}


\bibitem[Institute(2016)]%
        {reuters_institute_resources_2016}
\bibfield{author}{\bibinfo{person}{Reuters Institute}.} \bibinfo{year}{2016}\natexlab{}.
\newblock \bibinfo{title}{Resources and {Charts} for the 2016 {Digital} {News} {Report}}.
\newblock
\newblock
\urldef\tempurl%
\url{https://www.digitalnewsreport.org/survey/2016/resources-2016/}
\showURL{%
\tempurl}


\bibitem[Jhaver et~al\mbox{.}(2019)]%
        {jhaver_did_2019}
\bibfield{author}{\bibinfo{person}{Shagun Jhaver}, \bibinfo{person}{Darren~Scott Appling}, \bibinfo{person}{Eric Gilbert}, {and} \bibinfo{person}{Amy Bruckman}.} \bibinfo{year}{2019}\natexlab{}.
\newblock \showarticletitle{"{Did} {You} {Suspect} the {Post} {Would} be {Removed}?": {Understanding} {User} {Reactions} to {Content} {Removals} on {Reddit}}.
\newblock \bibinfo{journal}{\emph{Proceedings of the ACM on Human-Computer Interaction}} \bibinfo{volume}{3}, \bibinfo{number}{CSCW} (\bibinfo{date}{Nov.} \bibinfo{year}{2019}), \bibinfo{pages}{192:1--192:33}.
\newblock
\urldef\tempurl%
\url{https://doi.org/10.1145/3359294}
\showDOI{\tempurl}


\bibitem[Kammerer and Gerjets(2013)]%
        {kammerer_effects_2013}
\bibfield{author}{\bibinfo{person}{Yvonne Kammerer} {and} \bibinfo{person}{Peter Gerjets}.} \bibinfo{year}{2013}\natexlab{}.
\newblock \showarticletitle{"{Effects} of search interface and internet-specific epistemic beliefs on source evaluations during {Web} search for medical information: {An} eye-tracking study": {Corrigendum}}.
\newblock \bibinfo{journal}{\emph{Behaviour \& Information Technology}} \bibinfo{volume}{32}, \bibinfo{number}{7} (\bibinfo{year}{2013}), \bibinfo{pages}{747--747}.
\newblock
\showISSN{1362-3001}
\urldef\tempurl%
\url{https://doi.org/10.1080/0144929X.2011.633820}
\showDOI{\tempurl}
\newblock
\shownote{Place: United Kingdom Publisher: Taylor \& Francis}.


\bibitem[Kay et~al\mbox{.}(2015)]%
        {kay_unequal_2015}
\bibfield{author}{\bibinfo{person}{Matthew Kay}, \bibinfo{person}{Cynthia Matuszek}, {and} \bibinfo{person}{Sean~A. Munson}.} \bibinfo{year}{2015}\natexlab{}.
\newblock \showarticletitle{Unequal {Representation} and {Gender} {Stereotypes} in {Image} {Search} {Results} for {Occupations}}. In \bibinfo{booktitle}{\emph{Proceedings of the 33rd {Annual} {ACM} {Conference} on {Human} {Factors} in {Computing} {Systems}}} \emph{(\bibinfo{series}{{CHI} '15})}. \bibinfo{publisher}{Association for Computing Machinery}, \bibinfo{address}{New York, NY, USA}, \bibinfo{pages}{3819--3828}.
\newblock
\showISBNx{978-1-4503-3145-6}
\urldef\tempurl%
\url{https://doi.org/10.1145/2702123.2702520}
\showDOI{\tempurl}


\bibitem[Kim(2003)]%
        {kim_assessing_2003}
\bibfield{author}{\bibinfo{person}{Ji-Hyun Kim}.} \bibinfo{year}{2003}\natexlab{}.
\newblock \showarticletitle{Assessing practical significance of the proportional odds assumption}.
\newblock \bibinfo{journal}{\emph{Statistics \& Probability Letters}} \bibinfo{volume}{65}, \bibinfo{number}{3} (\bibinfo{date}{Nov.} \bibinfo{year}{2003}), \bibinfo{pages}{233--239}.
\newblock
\showISSN{0167-7152}
\urldef\tempurl%
\url{https://doi.org/10.1016/j.spl.2003.07.017}
\showDOI{\tempurl}


\bibitem[Knobloch-Westerwick et~al\mbox{.}(2015)]%
        {knobloch-westerwick_confirmation_2015}
\bibfield{author}{\bibinfo{person}{Silvia Knobloch-Westerwick}, \bibinfo{person}{Benjamin~K. Johnson}, {and} \bibinfo{person}{Axel Westerwick}.} \bibinfo{year}{2015}\natexlab{}.
\newblock \showarticletitle{Confirmation {Bias} in {Online} {Searches}: {Impacts} of {Selective} {Exposure} {Before} an {Election} on {Political} {Attitude} {Strength} and {Shifts}}.
\newblock \bibinfo{journal}{\emph{Journal of Computer-Mediated Communication}} \bibinfo{volume}{20}, \bibinfo{number}{2} (\bibinfo{date}{March} \bibinfo{year}{2015}), \bibinfo{pages}{171--187}.
\newblock
\showISSN{1083-6101}
\urldef\tempurl%
\url{https://doi.org/10.1111/jcc4.12105}
\showDOI{\tempurl}


\bibitem[Kozyreva et~al\mbox{.}(2022)]%
        {kozyreva_free_2022}
\bibfield{author}{\bibinfo{person}{Anastasia Kozyreva}, \bibinfo{person}{Stefan Herzog}, \bibinfo{person}{Stephan Lewandowsky}, \bibinfo{person}{Ralph Hertwig}, \bibinfo{person}{Philipp Lorenz-Spreen}, \bibinfo{person}{Mark Leiser}, {and} \bibinfo{person}{Jason Reifler}.} \bibinfo{year}{2022}\natexlab{}.
\newblock \bibinfo{title}{Free speech vs. harmful misinformation: {Moral} dilemmas in online content moderation}.
\newblock
\newblock
\urldef\tempurl%
\url{https://doi.org/10.31234/osf.io/2pc3a}
\showDOI{\tempurl}


\bibitem[Kroh(2007)]%
        {kroh_measuring_2007}
\bibfield{author}{\bibinfo{person}{Martin Kroh}.} \bibinfo{year}{2007}\natexlab{}.
\newblock \showarticletitle{Measuring {Left}-{Right} {Political} {Orientation}: {The} {Choice} of {Response} {Format}}.
\newblock \bibinfo{journal}{\emph{The Public Opinion Quarterly}} \bibinfo{volume}{71}, \bibinfo{number}{2} (\bibinfo{year}{2007}), \bibinfo{pages}{204--220}.
\newblock
\showISSN{0033-362X}
\urldef\tempurl%
\url{https://www.jstor.org/stable/4500371}
\showURL{%
\tempurl}
\newblock
\shownote{Publisher: [Oxford University Press, American Association for Public Opinion Research]}.


\bibitem[Lomas(2022)]%
        {lomas_russian_2022}
\bibfield{author}{\bibinfo{person}{Natasha Lomas}.} \bibinfo{year}{2022}\natexlab{}.
\newblock \bibinfo{title}{Russian tech giant {Yandex} removes national borders from {Maps} app}.
\newblock
\newblock
\urldef\tempurl%
\url{https://techcrunch.com/2022/06/09/yandex-maps-no-borders/}
\showURL{%
\tempurl}


\bibitem[Makhortykh et~al\mbox{.}(2021)]%
        {makhortykh_hey_2021}
\bibfield{author}{\bibinfo{person}{Mykola Makhortykh}, \bibinfo{person}{Aleksandra Urman}, {and} \bibinfo{person}{Roberto Ulloa}.} \bibinfo{year}{2021}\natexlab{}.
\newblock \showarticletitle{Hey, {Google}, is it what the {Holocaust} looked like?: {Auditing} algorithmic curation of visual historical content on {Web} search engines}.
\newblock \bibinfo{journal}{\emph{First Monday}} (\bibinfo{date}{Oct.} \bibinfo{year}{2021}).
\newblock
\showISSN{1396-0466}
\urldef\tempurl%
\url{https://doi.org/10.5210/fm.v26i10.11562}
\showDOI{\tempurl}


\bibitem[Makhortykh et~al\mbox{.}(2022a)]%
        {makhortykh_memory_2022}
\bibfield{author}{\bibinfo{person}{Mykola Makhortykh}, \bibinfo{person}{Aleksandra Urman}, {and} \bibinfo{person}{Roberto Ulloa}.} \bibinfo{year}{2022}\natexlab{a}.
\newblock \showarticletitle{Memory, counter-memory and denialism: {How} search engines circulate information about the {Holodomor}-related memory wars}.
\newblock \bibinfo{journal}{\emph{Memory Studies}} \bibinfo{volume}{15}, \bibinfo{number}{6} (\bibinfo{date}{Dec.} \bibinfo{year}{2022}), \bibinfo{pages}{1330--1345}.
\newblock
\showISSN{1750-6980}
\urldef\tempurl%
\url{https://doi.org/10.1177/17506980221133732}
\showDOI{\tempurl}
\newblock
\shownote{Publisher: SAGE Publications}.


\bibitem[Makhortykh et~al\mbox{.}(2022b)]%
        {makhortykh_story_2022}
\bibfield{author}{\bibinfo{person}{Mykola Makhortykh}, \bibinfo{person}{Aleksandra Urman}, {and} \bibinfo{person}{Mariëlle Wijermars}.} \bibinfo{year}{2022}\natexlab{b}.
\newblock \showarticletitle{A story of (non)compliance, bias, and conspiracies: {How} {Google} and {Yandex} represented {Smart} {Voting} during the 2021 parliamentary elections in {Russia}}.
\newblock \bibinfo{journal}{\emph{Harvard Kennedy School Misinformation Review}} (\bibinfo{date}{March} \bibinfo{year}{2022}).
\newblock
\urldef\tempurl%
\url{https://doi.org/10.37016/mr-2020-94}
\showDOI{\tempurl}


\bibitem[Meisner et~al\mbox{.}(2022)]%
        {meisner_labor_2022}
\bibfield{author}{\bibinfo{person}{Colten Meisner}, \bibinfo{person}{Brooke~Erin Duffy}, {and} \bibinfo{person}{Malte Ziewitz}.} \bibinfo{year}{2022}\natexlab{}.
\newblock \showarticletitle{The labor of search engine evaluation: {Making} algorithms more human or humans more algorithmic?}
\newblock \bibinfo{journal}{\emph{New Media \& Society}} (\bibinfo{date}{Jan.} \bibinfo{year}{2022}), \bibinfo{pages}{14614448211063860}.
\newblock
\showISSN{1461-4448}
\urldef\tempurl%
\url{https://doi.org/10.1177/14614448211063860}
\showDOI{\tempurl}
\newblock
\shownote{Publisher: SAGE Publications}.


\bibitem[Metaxa et~al\mbox{.}(2021)]%
        {metaxa_image_2021}
\bibfield{author}{\bibinfo{person}{Danaë Metaxa}, \bibinfo{person}{Michelle~A. Gan}, \bibinfo{person}{Su Goh}, \bibinfo{person}{Jeff Hancock}, {and} \bibinfo{person}{James~A. Landay}.} \bibinfo{year}{2021}\natexlab{}.
\newblock \showarticletitle{An {Image} of {Society}: {Gender} and {Racial} {Representation} and {Impact} in {Image} {Search} {Results} for {Occupations}}.
\newblock \bibinfo{journal}{\emph{Proceedings of the ACM on Human-Computer Interaction}} \bibinfo{volume}{5}, \bibinfo{number}{CSCW1} (\bibinfo{date}{April} \bibinfo{year}{2021}), \bibinfo{pages}{26:1--26:23}.
\newblock
\urldef\tempurl%
\url{https://doi.org/10.1145/3449100}
\showDOI{\tempurl}


\bibitem[Microsoft(2022)]%
        {microsoft_how_nodate}
\bibfield{author}{\bibinfo{person}{Microsoft}.} \bibinfo{year}{2022}\natexlab{}.
\newblock \bibinfo{title}{How {Bing} delivers search results - {Microsoft} {Support}}.
\newblock
\newblock
\urldef\tempurl%
\url{https://support.microsoft.com/en-us/topic/how-bing-delivers-search-results-d18fc815-ac37-4723-bc67-9229ce3eb6a3}
\showURL{%
\tempurl}


\bibitem[Molina and Sundar(2022)]%
        {molina_does_2022}
\bibfield{author}{\bibinfo{person}{Maria~D. Molina} {and} \bibinfo{person}{S.~Shyam Sundar}.} \bibinfo{year}{2022}\natexlab{}.
\newblock \showarticletitle{Does distrust in humans predict greater trust in {AI}? {Role} of individual differences in user responses to content moderation}.
\newblock \bibinfo{journal}{\emph{New Media \& Society}} (\bibinfo{date}{June} \bibinfo{year}{2022}), \bibinfo{pages}{14614448221103534}.
\newblock
\showISSN{1461-4448}
\urldef\tempurl%
\url{https://doi.org/10.1177/14614448221103534}
\showDOI{\tempurl}
\newblock
\shownote{Publisher: SAGE Publications}.


\bibitem[Morrow et~al\mbox{.}(2022)]%
        {morrow_emerging_2022}
\bibfield{author}{\bibinfo{person}{Garrett Morrow}, \bibinfo{person}{Briony Swire-Thompson}, \bibinfo{person}{Jessica~Montgomery Polny}, \bibinfo{person}{Matthew Kopec}, {and} \bibinfo{person}{John~P. Wihbey}.} \bibinfo{year}{2022}\natexlab{}.
\newblock \showarticletitle{The emerging science of content labeling: {Contextualizing} social media content moderation}.
\newblock \bibinfo{journal}{\emph{Journal of the Association for Information Science and Technology}} \bibinfo{volume}{73}, \bibinfo{number}{10} (\bibinfo{year}{2022}), \bibinfo{pages}{1365--1386}.
\newblock
\showISSN{2330-1643}
\urldef\tempurl%
\url{https://doi.org/10.1002/asi.24637}
\showDOI{\tempurl}
\newblock
\shownote{\_eprint: https://onlinelibrary.wiley.com/doi/pdf/10.1002/asi.24637}.


\bibitem[Myers~West(2018)]%
        {myers_west_censored_2018}
\bibfield{author}{\bibinfo{person}{Sarah Myers~West}.} \bibinfo{year}{2018}\natexlab{}.
\newblock \showarticletitle{Censored, suspended, shadowbanned: {User} interpretations of content moderation on social media platforms}.
\newblock \bibinfo{journal}{\emph{New Media \& Society}} \bibinfo{volume}{20}, \bibinfo{number}{11} (\bibinfo{date}{Nov.} \bibinfo{year}{2018}), \bibinfo{pages}{4366--4383}.
\newblock
\showISSN{1461-4448}
\urldef\tempurl%
\url{https://doi.org/10.1177/1461444818773059}
\showDOI{\tempurl}
\newblock
\shownote{Publisher: SAGE Publications}.


\bibitem[Noble(2018)]%
        {noble_algorithms_2018}
\bibfield{author}{\bibinfo{person}{Safiya~Umoja Noble}.} \bibinfo{year}{2018}\natexlab{}.
\newblock \bibinfo{booktitle}{\emph{Algorithms of {Oppression}: {How} {Search} {Engines} {Reinforce} {Racism}}}.
\newblock \bibinfo{publisher}{New York University Press}.
\newblock
\showISBNx{978-1-4798-3364-1}
\urldef\tempurl%
\url{https://doi.org/10.18574/9781479833641}
\showDOI{\tempurl}
\newblock
\shownote{Publication Title: Algorithms of Oppression}.


\bibitem[Ozanne et~al\mbox{.}(2022)]%
        {ozanne_shall_2022}
\bibfield{author}{\bibinfo{person}{Marie Ozanne}, \bibinfo{person}{Aparajita Bhandari}, \bibinfo{person}{Natalya~N Bazarova}, {and} \bibinfo{person}{Dominic DiFranzo}.} \bibinfo{year}{2022}\natexlab{}.
\newblock \showarticletitle{Shall {AI} moderators be made visible? {Perception} of accountability and trust in moderation systems on social media platforms}.
\newblock \bibinfo{journal}{\emph{Big Data \& Society}} \bibinfo{volume}{9}, \bibinfo{number}{2} (\bibinfo{date}{July} \bibinfo{year}{2022}), \bibinfo{pages}{205395172211156}.
\newblock
\showISSN{2053-9517, 2053-9517}
\urldef\tempurl%
\url{https://doi.org/10.1177/20539517221115666}
\showDOI{\tempurl}


\bibitem[Pan et~al\mbox{.}(2007)]%
        {pan_google_2007}
\bibfield{author}{\bibinfo{person}{Bing Pan}, \bibinfo{person}{Helene Hembrooke}, \bibinfo{person}{Thorsten Joachims}, \bibinfo{person}{Lori Lorigo}, \bibinfo{person}{Geri Gay}, {and} \bibinfo{person}{Laura Granka}.} \bibinfo{year}{2007}\natexlab{}.
\newblock \showarticletitle{In {Google} {We} {Trust}: {Users}’ {Decisions} on {Rank}, {Position}, and {Relevance}}.
\newblock \bibinfo{journal}{\emph{Journal of Computer-Mediated Communication}} \bibinfo{volume}{12}, \bibinfo{number}{3} (\bibinfo{date}{April} \bibinfo{year}{2007}), \bibinfo{pages}{801--823}.
\newblock
\showISSN{1083-6101}
\urldef\tempurl%
\url{https://doi.org/10.1111/j.1083-6101.2007.00351.x}
\showDOI{\tempurl}


\bibitem[Pan et~al\mbox{.}(2022)]%
        {pan_comparing_2022}
\bibfield{author}{\bibinfo{person}{Christina~A. Pan}, \bibinfo{person}{Sahil Yakhmi}, \bibinfo{person}{Tara~P. Iyer}, \bibinfo{person}{Evan Strasnick}, \bibinfo{person}{Amy~X. Zhang}, {and} \bibinfo{person}{Michael~S. Bernstein}.} \bibinfo{year}{2022}\natexlab{}.
\newblock \showarticletitle{Comparing the {Perceived} {Legitimacy} of {Content} {Moderation} {Processes}: {Contractors}, {Algorithms}, {Expert} {Panels}, and {Digital} {Juries}}.
\newblock \bibinfo{journal}{\emph{Proceedings of the ACM on Human-Computer Interaction}} \bibinfo{volume}{6}, \bibinfo{number}{CSCW1} (\bibinfo{date}{April} \bibinfo{year}{2022}), \bibinfo{pages}{82:1--82:31}.
\newblock
\urldef\tempurl%
\url{https://doi.org/10.1145/3512929}
\showDOI{\tempurl}


\bibitem[Peterson and Harrell(1990)]%
        {peterson_partial_1990}
\bibfield{author}{\bibinfo{person}{Bercedis Peterson} {and} \bibinfo{person}{Frank~E. Harrell}.} \bibinfo{year}{1990}\natexlab{}.
\newblock \showarticletitle{Partial {Proportional} {Odds} {Models} for {Ordinal} {Response} {Variables}}.
\newblock \bibinfo{journal}{\emph{Applied Statistics}} \bibinfo{volume}{39}, \bibinfo{number}{2} (\bibinfo{year}{1990}), \bibinfo{pages}{205}.
\newblock
\showISSN{00359254}
\urldef\tempurl%
\url{https://doi.org/10.2307/2347760}
\showDOI{\tempurl}


\bibitem[Pradel et~al\mbox{.}(2022)]%
        {pradel_users_2022}
\bibfield{author}{\bibinfo{person}{Franziska Pradel}, \bibinfo{person}{Jan Zilinsky}, \bibinfo{person}{Spyros Kosmidis}, {and} \bibinfo{person}{Yannis Theocharis}.} \bibinfo{year}{2022}\natexlab{}.
\newblock \bibinfo{title}{Do {Users} {Ever} {Draw} a {Line}? {Offensiveness} and {Content} {Moderation} {Preferences} on {Social} {Media}}.
\newblock
\newblock
\urldef\tempurl%
\url{https://doi.org/10.31219/osf.io/y4xft}
\showDOI{\tempurl}


\bibitem[Prolific(2022)]%
        {prolific_representative_2022}
\bibfield{author}{\bibinfo{person}{Prolific}.} \bibinfo{year}{2022}\natexlab{}.
\newblock \bibinfo{title}{Representative samples}.
\newblock
\newblock
\urldef\tempurl%
\url{https://researcher-help.prolific.co/hc/en-gb/articles/360019236753-Representative-samples}
\showURL{%
\tempurl}


\bibitem[Riedl et~al\mbox{.}(2022)]%
        {riedl_antecedents_2022}
\bibfield{author}{\bibinfo{person}{Martin~J. Riedl}, \bibinfo{person}{Kelsey~N. Whipple}, {and} \bibinfo{person}{Ryan Wallace}.} \bibinfo{year}{2022}\natexlab{}.
\newblock \showarticletitle{Antecedents of support for social media content moderation and platform regulation: the role of presumed effects on self and others}.
\newblock \bibinfo{journal}{\emph{Information, Communication \& Society}} \bibinfo{volume}{25}, \bibinfo{number}{11} (\bibinfo{date}{Aug.} \bibinfo{year}{2022}), \bibinfo{pages}{1632--1649}.
\newblock
\showISSN{1369-118X}
\urldef\tempurl%
\url{https://doi.org/10.1080/1369118X.2021.1874040}
\showDOI{\tempurl}
\newblock
\shownote{Publisher: Routledge \_eprint: https://doi.org/10.1080/1369118X.2021.1874040}.


\bibitem[Ruokolainen and Widén(2020)]%
        {RUOKOLAINEN2020102127}
\bibfield{author}{\bibinfo{person}{Hilda Ruokolainen} {and} \bibinfo{person}{Gunilla Widén}.} \bibinfo{year}{2020}\natexlab{}.
\newblock \showarticletitle{Conceptualising misinformation in the context of asylum seekers}.
\newblock \bibinfo{journal}{\emph{Information Processing \& Management}} \bibinfo{volume}{57}, \bibinfo{number}{3} (\bibinfo{year}{2020}), \bibinfo{pages}{102127}.
\newblock
\showISSN{0306-4573}
\urldef\tempurl%
\url{https://doi.org/10.1016/j.ipm.2019.102127}
\showDOI{\tempurl}


\bibitem[Saltz et~al\mbox{.}(2021)]%
        {saltz_encounters_2021}
\bibfield{author}{\bibinfo{person}{Emily Saltz}, \bibinfo{person}{Claire~R Leibowicz}, {and} \bibinfo{person}{Claire Wardle}.} \bibinfo{year}{2021}\natexlab{}.
\newblock \showarticletitle{Encounters with {Visual} {Misinformation} and {Labels} {Across} {Platforms}: {An} {Interview} and {Diary} {Study} to {Inform} {Ecosystem} {Approaches} to {Misinformation} {Interventions}}. In \bibinfo{booktitle}{\emph{Extended {Abstracts} of the 2021 {CHI} {Conference} on {Human} {Factors} in {Computing} {Systems}}} \emph{(\bibinfo{series}{{CHI} {EA} '21})}. \bibinfo{publisher}{Association for Computing Machinery}, \bibinfo{address}{New York, NY, USA}, \bibinfo{pages}{1--6}.
\newblock
\showISBNx{978-1-4503-8095-9}
\urldef\tempurl%
\url{https://doi.org/10.1145/3411763.3451807}
\showDOI{\tempurl}


\bibitem[Scherr et~al\mbox{.}(2022)]%
        {scherr_algorithms_2022}
\bibfield{author}{\bibinfo{person}{Sebastian Scherr}, \bibinfo{person}{Florian Arendt}, {and} \bibinfo{person}{Mario Haim}.} \bibinfo{year}{2022}\natexlab{}.
\newblock \showarticletitle{Algorithms without frontiers? {How} language-based algorithmic information disparities for suicide crisis information sustain digital divides over time in 17 countries}.
\newblock \bibinfo{journal}{\emph{Information, Communication \& Society}} \bibinfo{volume}{0}, \bibinfo{number}{0} (\bibinfo{date}{July} \bibinfo{year}{2022}), \bibinfo{pages}{1--17}.
\newblock
\showISSN{1369-118X}
\urldef\tempurl%
\url{https://doi.org/10.1080/1369118X.2022.2097017}
\showDOI{\tempurl}
\newblock
\shownote{Publisher: Routledge \_eprint: https://doi.org/10.1080/1369118X.2022.2097017}.


\bibitem[Scherr et~al\mbox{.}(2019)]%
        {scherr_equal_2019}
\bibfield{author}{\bibinfo{person}{Sebastian Scherr}, \bibinfo{person}{Mario Haim}, {and} \bibinfo{person}{Florian Arendt}.} \bibinfo{year}{2019}\natexlab{}.
\newblock \showarticletitle{Equal access to online information? {Google}’s suicide-prevention disparities may amplify a global digital divide}.
\newblock \bibinfo{journal}{\emph{New Media \& Society}} \bibinfo{volume}{21}, \bibinfo{number}{3} (\bibinfo{date}{March} \bibinfo{year}{2019}), \bibinfo{pages}{562--582}.
\newblock
\showISSN{1461-4448}
\urldef\tempurl%
\url{https://doi.org/10.1177/1461444818801010}
\showDOI{\tempurl}
\newblock
\shownote{Publisher: SAGE Publications}.


\bibitem[Schoenebeck et~al\mbox{.}(2021)]%
        {schoenebeck_youth_2021}
\bibfield{author}{\bibinfo{person}{Sarita Schoenebeck}, \bibinfo{person}{Carol~F. Scott}, \bibinfo{person}{Emma~Grace Hurley}, \bibinfo{person}{Tammy Chang}, {and} \bibinfo{person}{Ellen Selkie}.} \bibinfo{year}{2021}\natexlab{}.
\newblock \showarticletitle{Youth {Trust} in {Social} {Media} {Companies} and {Expectations} of {Justice}: {Accountability} and {Repair} {After} {Online} {Harassment}}.
\newblock \bibinfo{journal}{\emph{Proceedings of the ACM on Human-Computer Interaction}} \bibinfo{volume}{5}, \bibinfo{number}{CSCW1} (\bibinfo{date}{April} \bibinfo{year}{2021}), \bibinfo{pages}{2:1--2:18}.
\newblock
\urldef\tempurl%
\url{https://doi.org/10.1145/3449076}
\showDOI{\tempurl}


\bibitem[Schultheiß et~al\mbox{.}(2018)]%
        {schultheis_we_2018}
\bibfield{author}{\bibinfo{person}{Sebastian Schultheiß}, \bibinfo{person}{Sebastian Sünkler}, {and} \bibinfo{person}{Dirk Lewandowski}.} \bibinfo{year}{2018}\natexlab{}.
\newblock \showarticletitle{We {Still} {Trust} in {Google}, but {Less} than 10 {Years} {Ago}: {An} {Eye}-{Tracking} {Study}}.
\newblock \bibinfo{journal}{\emph{Information Research: An International Electronic Journal}} \bibinfo{volume}{23}, \bibinfo{number}{3} (\bibinfo{date}{Sept.} \bibinfo{year}{2018}).
\newblock
\showISSN{1368-1613}
\urldef\tempurl%
\url{https://eric.ed.gov/?id=EJ1196314}
\showURL{%
\tempurl}
\newblock
\shownote{Publisher: Thomas D}.


\bibitem[Statcounter(2022)]%
        {statcounter_search_2022}
\bibfield{author}{\bibinfo{person}{Statcounter}.} \bibinfo{year}{2022}\natexlab{}.
\newblock \bibinfo{title}{Search {Engine} {Market} {Share} {Worldwide}}.
\newblock
\newblock
\urldef\tempurl%
\url{https://gs.statcounter.com/search-engine-market-share}
\showURL{%
\tempurl}


\bibitem[Strömbäck et~al\mbox{.}(2020)]%
        {stromback_news_2020}
\bibfield{author}{\bibinfo{person}{Jesper Strömbäck}, \bibinfo{person}{Yariv Tsfati}, \bibinfo{person}{Hajo Boomgaarden}, \bibinfo{person}{Alyt Damstra}, \bibinfo{person}{Elina Lindgren}, \bibinfo{person}{Rens Vliegenthart}, {and} \bibinfo{person}{Torun Lindholm}.} \bibinfo{year}{2020}\natexlab{}.
\newblock \showarticletitle{News media trust and its impact on media use: toward a framework for future research}.
\newblock \bibinfo{journal}{\emph{Annals of the International Communication Association}} \bibinfo{volume}{44}, \bibinfo{number}{2} (\bibinfo{date}{April} \bibinfo{year}{2020}), \bibinfo{pages}{139--156}.
\newblock
\showISSN{2380-8985}
\urldef\tempurl%
\url{https://doi.org/10.1080/23808985.2020.1755338}
\showDOI{\tempurl}
\newblock
\shownote{Publisher: Routledge \_eprint: https://doi.org/10.1080/23808985.2020.1755338}.


\bibitem[Thompson et~al\mbox{.}(2017)]%
        {thompson_extracting_2017}
\bibfield{author}{\bibinfo{person}{Christopher~Glen Thompson}, \bibinfo{person}{Rae~Seon Kim}, \bibinfo{person}{Ariel~M. Aloe}, {and} \bibinfo{person}{Betsy~Jane Becker}.} \bibinfo{year}{2017}\natexlab{}.
\newblock \showarticletitle{Extracting the {Variance} {Inflation} {Factor} and {Other} {Multicollinearity} {Diagnostics} from {Typical} {Regression} {Results}}.
\newblock \bibinfo{journal}{\emph{Basic and Applied Social Psychology}} \bibinfo{volume}{39}, \bibinfo{number}{2} (\bibinfo{date}{March} \bibinfo{year}{2017}), \bibinfo{pages}{81--90}.
\newblock
\showISSN{0197-3533}
\urldef\tempurl%
\url{https://doi.org/10.1080/01973533.2016.1277529}
\showDOI{\tempurl}
\newblock
\shownote{Publisher: Routledge \_eprint: https://doi.org/10.1080/01973533.2016.1277529}.


\bibitem[Thompson(2022)]%
        {thompson_fed_2022}
\bibfield{author}{\bibinfo{person}{Stuart~A. Thompson}.} \bibinfo{year}{2022}\natexlab{}.
\newblock \showarticletitle{Fed {Up} {With} {Google}, {Conspiracy} {Theorists} {Turn} to {DuckDuckGo}}.
\newblock \bibinfo{journal}{\emph{The New York Times}} (\bibinfo{date}{Feb.} \bibinfo{year}{2022}).
\newblock
\showISSN{0362-4331}
\urldef\tempurl%
\url{https://www.nytimes.com/2022/02/23/technology/duckduckgo-conspiracy-theories.html}
\showURL{%
\tempurl}


\bibitem[Toepfl et~al\mbox{.}(2022)]%
        {toepfl_who_2022}
\bibfield{author}{\bibinfo{person}{Florian Toepfl}, \bibinfo{person}{Daria Kravets}, \bibinfo{person}{Anna Ryzhova}, {and} \bibinfo{person}{Arista Beseler}.} \bibinfo{year}{2022}\natexlab{}.
\newblock \showarticletitle{Who are the plotters behind the pandemic? {Comparing} {Covid}-19 conspiracy theories in {Google} search results across five key target countries of {Russia}’s foreign communication}.
\newblock \bibinfo{journal}{\emph{Information, Communication \& Society}} \bibinfo{volume}{0}, \bibinfo{number}{0} (\bibinfo{date}{April} \bibinfo{year}{2022}), \bibinfo{pages}{1--19}.
\newblock
\showISSN{1369-118X}
\urldef\tempurl%
\url{https://doi.org/10.1080/1369118X.2022.2065213}
\showDOI{\tempurl}
\newblock
\shownote{Publisher: Routledge \_eprint: https://doi.org/10.1080/1369118X.2022.2065213}.


\bibitem[Tripodi(2022a)]%
        {tripodi_searching_nodate}
\bibfield{author}{\bibinfo{person}{Francesca Tripodi}.} \bibinfo{year}{2022}\natexlab{a}.
\newblock \showarticletitle{Searching for {Alternative} {Facts}}.
\newblock \bibinfo{journal}{\emph{Data \& Society}} (\bibinfo{year}{2022}).
\newblock


\bibitem[Tripodi(2022b)]%
        {tripodi_propagandists_2022}
\bibfield{author}{\bibinfo{person}{Francesca~Bolla Tripodi}.} \bibinfo{year}{2022}\natexlab{b}.
\newblock \bibinfo{booktitle}{\emph{The propagandists' playbook: how conservative elites manipulate search and threaten democracy}}.
\newblock \bibinfo{publisher}{Yale University Press}, \bibinfo{address}{New Haven}.
\newblock
\showISBNx{978-0-300-24894-4}
\newblock
\shownote{OCLC: on1305434359}.


\bibitem[Ulloa et~al\mbox{.}(2022)]%
        {ulloa_representativeness_2022}
\bibfield{author}{\bibinfo{person}{Roberto Ulloa}, \bibinfo{person}{Ana~Carolina Richter}, \bibinfo{person}{Mykola Makhortykh}, \bibinfo{person}{Aleksandra Urman}, {and} \bibinfo{person}{Celina~Sylwia Kacperski}.} \bibinfo{year}{2022}\natexlab{}.
\newblock \showarticletitle{Representativeness and face-ism: {Gender} bias in image search}.
\newblock \bibinfo{journal}{\emph{New Media \& Society}} (\bibinfo{date}{June} \bibinfo{year}{2022}), \bibinfo{pages}{14614448221100699}.
\newblock
\showISSN{1461-4448}
\urldef\tempurl%
\url{https://doi.org/10.1177/14614448221100699}
\showDOI{\tempurl}
\newblock
\shownote{Publisher: SAGE Publications}.


\bibitem[Urman and Makhortykh(2021)]%
        {urman_you_2021}
\bibfield{author}{\bibinfo{person}{Aleksandra Urman} {and} \bibinfo{person}{Mykola Makhortykh}.} \bibinfo{year}{2021}\natexlab{}.
\newblock \bibinfo{title}{You {Are} {How} (and {Where}) {You} {Search}? {Comparative} {Analysis} of {Web} {Search} {Behaviour} {Using} {Web} {Tracking} {Data}}.
\newblock
\newblock
\urldef\tempurl%
\url{https://doi.org/10.48550/arXiv.2105.04961}
\showDOI{\tempurl}
\newblock
\shownote{arXiv:2105.04961 [cs]}.


\bibitem[Urman and Makhortykh(2022)]%
        {urman_foreign_2022}
\bibfield{author}{\bibinfo{person}{Aleksandra Urman} {and} \bibinfo{person}{Mykola Makhortykh}.} \bibinfo{year}{2022}\natexlab{}.
\newblock \showarticletitle{“{Foreign} beauties want to meet you”: {The} sexualization of women in {Google}’s organic and sponsored text search results}.
\newblock \bibinfo{journal}{\emph{New Media \& Society}} (\bibinfo{date}{June} \bibinfo{year}{2022}), \bibinfo{pages}{14614448221099536}.
\newblock
\showISSN{1461-4448}
\urldef\tempurl%
\url{https://doi.org/10.1177/14614448221099536}
\showDOI{\tempurl}
\newblock
\shownote{Publisher: SAGE Publications}.


\bibitem[Urman and Makhortykh(2023)]%
        {urman_how_2023}
\bibfield{author}{\bibinfo{person}{Aleksandra Urman} {and} \bibinfo{person}{Mykola Makhortykh}.} \bibinfo{year}{2023}\natexlab{}.
\newblock \showarticletitle{How transparent are transparency reports? {Comparative} analysis of transparency reporting across online platforms}.
\newblock \bibinfo{journal}{\emph{Telecommunications Policy}} (\bibinfo{date}{Jan.} \bibinfo{year}{2023}), \bibinfo{pages}{102477}.
\newblock
\showISSN{03085961}
\urldef\tempurl%
\url{https://doi.org/10.1016/j.telpol.2022.102477}
\showDOI{\tempurl}


\bibitem[Urman et~al\mbox{.}(2022)]%
        {urman_where_2022}
\bibfield{author}{\bibinfo{person}{Aleksandra Urman}, \bibinfo{person}{Mykola Makhortykh}, \bibinfo{person}{Roberto Ulloa}, {and} \bibinfo{person}{Juhi Kulshrestha}.} \bibinfo{year}{2022}\natexlab{}.
\newblock \showarticletitle{Where the earth is flat and 9/11 is an inside job: {A} comparative algorithm audit of conspiratorial information in web search results}.
\newblock \bibinfo{journal}{\emph{Telematics and Informatics}}  \bibinfo{volume}{72} (\bibinfo{date}{Aug.} \bibinfo{year}{2022}), \bibinfo{pages}{101860}.
\newblock
\showISSN{0736-5853}
\urldef\tempurl%
\url{https://doi.org/10.1016/j.tele.2022.101860}
\showDOI{\tempurl}


\bibitem[Vaccaro et~al\mbox{.}(2020)]%
        {vaccaro_at_2020}
\bibfield{author}{\bibinfo{person}{Kristen Vaccaro}, \bibinfo{person}{Christian Sandvig}, {and} \bibinfo{person}{Karrie Karahalios}.} \bibinfo{year}{2020}\natexlab{}.
\newblock \showarticletitle{"{At} the {End} of the {Day} {Facebook} {Does} {What} {ItWants}": {How} {Users} {Experience} {Contesting} {Algorithmic} {Content} {Moderation}}.
\newblock \bibinfo{journal}{\emph{Proceedings of the ACM on Human-Computer Interaction}} \bibinfo{volume}{4}, \bibinfo{number}{CSCW2} (\bibinfo{date}{Oct.} \bibinfo{year}{2020}), \bibinfo{pages}{1--22}.
\newblock
\showISSN{2573-0142}
\urldef\tempurl%
\url{https://doi.org/10.1145/3415238}
\showDOI{\tempurl}


\bibitem[Vlasceanu and Amodio(2022)]%
        {vlasceanu_propagation_2022}
\bibfield{author}{\bibinfo{person}{Madalina Vlasceanu} {and} \bibinfo{person}{David~M. Amodio}.} \bibinfo{year}{2022}\natexlab{}.
\newblock \showarticletitle{Propagation of societal gender inequality by internet search algorithms}.
\newblock \bibinfo{journal}{\emph{Proceedings of the National Academy of Sciences}} \bibinfo{volume}{119}, \bibinfo{number}{29} (\bibinfo{date}{July} \bibinfo{year}{2022}), \bibinfo{pages}{e2204529119}.
\newblock
\urldef\tempurl%
\url{https://doi.org/10.1073/pnas.2204529119}
\showDOI{\tempurl}
\newblock
\shownote{Publisher: Proceedings of the National Academy of Sciences}.


\bibitem[Wihbey et~al\mbox{.}(2021)]%
        {wihbey_bipartisan_2021}
\bibfield{author}{\bibinfo{person}{John Wihbey}, \bibinfo{person}{Garrett Morrow}, \bibinfo{person}{Myojung Chung}, {and} \bibinfo{person}{Mike Peacey}.} \bibinfo{year}{2021}\natexlab{}.
\newblock \bibinfo{title}{The {Bipartisan} {Case} for {Labeling} as a {Content} {Moderation} {Method}: {Findings} from a {National} {Survey}}.
\newblock
\newblock
\urldef\tempurl%
\url{https://doi.org/10.2139/ssrn.3923905}
\showDOI{\tempurl}


\bibitem[Xu et~al\mbox{.}(2021)]%
        {xu_how_2021}
\bibfield{author}{\bibinfo{person}{Luyan Xu}, \bibinfo{person}{Mengdie Zhuang}, {and} \bibinfo{person}{Ujwal Gadiraju}.} \bibinfo{year}{2021}\natexlab{}.
\newblock \showarticletitle{How {Do} {User} {Opinions} {Influence} {Their} {Interaction} {With} {Web} {Search} {Results}?}. In \bibinfo{booktitle}{\emph{Proceedings of the 29th {ACM} {Conference} on {User} {Modeling}, {Adaptation} and {Personalization}}} \emph{(\bibinfo{series}{{UMAP} '21})}. \bibinfo{publisher}{Association for Computing Machinery}, \bibinfo{address}{New York, NY, USA}, \bibinfo{pages}{240--244}.
\newblock
\showISBNx{978-1-4503-8366-0}
\urldef\tempurl%
\url{https://doi.org/10.1145/3450613.3456824}
\showDOI{\tempurl}


\bibitem[Yahoo!(2022a)]%
        {yahoo_remove_2022}
\bibfield{author}{\bibinfo{person}{Yahoo!}} \bibinfo{year}{2022}\natexlab{a}.
\newblock \bibinfo{title}{Remove search results from {Yahoo} {Search} {\textbar} {Yahoo} {Help} - {SLN4530}}.
\newblock
\newblock
\urldef\tempurl%
\url{https://help.yahoo.com/kb/SLN4530.html}
\showURL{%
\tempurl}


\bibitem[Yahoo!(2022b)]%
        {yahoo_search_2022}
\bibfield{author}{\bibinfo{person}{Yahoo!}} \bibinfo{year}{2022}\natexlab{b}.
\newblock \bibinfo{title}{Search {Services} {\textbar} {Yahoo}}.
\newblock
\newblock
\urldef\tempurl%
\url{https://legal.yahoo.com/in/en/yahoo/privacy/products/searchservices/index.html#yahoosearch}
\showURL{%
\tempurl}


\bibitem[Yandex(2022a)]%
        {yandex_signs_nodate}
\bibfield{author}{\bibinfo{person}{Yandex}.} \bibinfo{year}{2022}\natexlab{a}.
\newblock \bibinfo{title}{Signs of a low-quality site - {Webmaster}. {Help}}.
\newblock
\newblock
\urldef\tempurl%
\url{https://yandex.com/support/webmaster/yandex-indexing/webmaster-advice.html}
\showURL{%
\tempurl}


\bibitem[Yandex(2022b)]%
        {yandex_how_nodate}
\bibfield{author}{\bibinfo{person}{Yandex}.} \bibinfo{year}{2022}\natexlab{b}.
\newblock \bibinfo{title}{Why are pages excluded from the search?}
\newblock
\newblock
\urldef\tempurl%
\url{https://yandex.com/support/webmaster/site-indexing/excluded-pages.html}
\showURL{%
\tempurl}


\bibitem[Zhang et~al\mbox{.}(2015)]%
        {zhang_quality_2015}
\bibfield{author}{\bibinfo{person}{Yan Zhang}, \bibinfo{person}{Yalin Sun}, {and} \bibinfo{person}{Bo Xie}.} \bibinfo{year}{2015}\natexlab{}.
\newblock \showarticletitle{Quality of health information for consumers on the web: {A} systematic review of indicators, criteria, tools, and evaluation results}.
\newblock \bibinfo{journal}{\emph{Journal of the Association for Information Science and Technology}} \bibinfo{volume}{66}, \bibinfo{number}{10} (\bibinfo{year}{2015}), \bibinfo{pages}{2071--2084}.
\newblock
\showISSN{2330-1643}
\urldef\tempurl%
\url{https://doi.org/10.1002/asi.23311}
\showDOI{\tempurl}
\newblock
\shownote{\_eprint: https://onlinelibrary.wiley.com/doi/pdf/10.1002/asi.23311}.


\end{thebibliography}


\appendix

\section{Appendix}
\subsection{Appendix 1: The introductory consent statement presented to the respondents}

Dear participant,
 
Before you start with this survey, it is important that you know what this research entails. Therefore, please read this information carefully. If you don’t understand something, you can e-mail us. We are happy to answer your questions. You can find our contact information at the end of this page.
 
\textbf{What is the goal of this research?}

We are examining people’s attitudes towards web search engines and other online platforms people use to inform themselves.
 
\textbf{What does the research consist of?}

The research consists of a questionnaire about your media use, web search use and opinions on web search and other information platforms. This questionnaire will take about 20 minutes.
 
\textbf{What happens to my data?}

This research is conducted under the responsibility of the [NAME OF THE UNIVERSITY IS BLINDED FOR REVIEW].

We guarantee the following:

    Your data will be kept completely confidential and anonymous. We will not ask for your name or any other identifying information, which means that your information cannot be linked to your identity.
    We use the research data only to address our research questions. (See: “What is the goal of the research?”)
    The results of the research will only be used in scientific articles, and the data will not be used for any commercial purposes.

\textbf{Do I have to participate in this research?}

Participating in this research is voluntary. If you do not participate, this does not have any consequences for you, except we will not be able to pay you full remuneration through Prolific if you do not participate in the survey. If you have already started, you can always decide to stop with the research. You don’t have to give a reason for this.
 
\textbf{Will I get paid for the participation?}

Yes, you will get paid through Prolific. Please make sure to enter your Prolific ID correctly and copy the completion code that we provide at the end of the survey!
 
\textbf{Contact information}

If you have questions about this research, you can contact the responsible researcher.
 
The responsible researcher: Dr. Aleksandra Urman, University of Zurich. email: urman@ifi.uzh.ch
 
Thank you for your participation.
 
If you would like to participate in this research, please choose “Yes” below. By choosing “yes”, you confirm that you have read and understood the information above and that you are voluntarily participating in the study based on the information received.

\subsection{Appendix 2: Additional regression models}

\begin{table}[!htbp]
\begin{center}
\begin{tabular}{l c c c c c c}
\hline
 & Mis: Inform & Mis: Reduce & Mis: Remove & Off: Inform & Off: Reduce & Off: Remove \\
\hline
Age              & $0.01$        & $-0.01$       & $0.01$        & $0.00$        & $-0.01$      & $0.00$        \\
                 & $(0.01)$      & $(0.01)$      & $(0.01)$      & $(0.01)$      & $(0.01)$     & $(0.01)$      \\
G - Non-binary      & $0.58$        & $0.60$        & $0.76$        & $1.02$        & $0.40$       & $0.26$        \\
                 & $(0.77)$      & $(0.65)$      & $(0.64)$      & $(0.76)$      & $(0.60)$     & $(0.63)$      \\
G - Woman      & $-0.08$       & $0.26$        & $0.18$        & $0.01$        & $0.42^{*}$   & $0.17$        \\
                 & $(0.22)$      & $(0.20)$      & $(0.20)$      & $(0.21)$      & $(0.20)$     & $(0.19)$      \\
Education       & $-0.06$       & $0.00$        & $-0.04$       & $0.01$        & $0.02$       & $0.11$        \\
                 & $(0.08)$      & $(0.07)$      & $(0.07)$      & $(0.08)$      & $(0.07)$     & $(0.07)$      \\
Race (White)  & $0.32$        & $-0.11$       & $-0.22$       & $0.25$        & $-0.21$      & $-0.50^{*}$   \\
                 & $(0.24)$      & $(0.23)$      & $(0.22)$      & $(0.24)$      & $(0.22)$     & $(0.22)$      \\
Trust in SE           & $0.55^{***}$  & $0.29^{*}$    & $0.15$        & $0.25^{*}$    & $0.03$       & $0.10$        \\
                 & $(0.13)$      & $(0.12)$      & $(0.12)$      & $(0.12)$      & $(0.11)$     & $(0.11)$      \\
SE independence            & $0.06$        & $0.22^{**}$   & $0.26^{***}$  & $0.14$        & $0.31^{***}$ & $0.31^{***}$  \\
                 & $(0.08)$      & $(0.07)$      & $(0.07)$      & $(0.08)$      & $(0.07)$     & $(0.07)$      \\
Political ideology & $-0.25^{***}$ & $-0.20^{***}$ & $-0.20^{***}$ & $-0.14^{***}$ & $-0.10^{**}$ & $-0.04$       \\
                 & $(0.04)$      & $(0.04)$      & $(0.04)$      & $(0.04)$      & $(0.04)$     & $(0.04)$      \\
Google use  & $0.18^{*}$    & $0.28^{***}$  & $0.17^{*}$    & $0.20^{*}$    & $0.13$       & $0.10$        \\
                 & $(0.09)$      & $(0.08)$      & $(0.08)$      & $(0.08)$      & $(0.08)$     & $(0.08)$      \\
DDG use    & $-0.06$       & $-0.09$       & $-0.14^{*}$   & $-0.08$       & $-0.14^{**}$ & $-0.19^{***}$ \\
                 & $(0.06)$      & $(0.05)$      & $(0.05)$      & $(0.06)$      & $(0.05)$     & $(0.05)$      \\
Yandex use   & $-0.12$       & $0.02$        & $0.00$        & $-0.20$       & $-0.02$      & $0.02$        \\
                 & $(0.14)$      & $(0.14)$      & $(0.15)$      & $(0.14)$      & $(0.14)$     & $(0.14)$      \\
Yahoo use   & $0.01$        & $0.06$        & $0.07$        & $0.03$        & $0.12$       & $0.13^{*}$    \\
                 & $(0.07)$      & $(0.07)$      & $(0.06)$      & $(0.07)$      & $(0.06)$     & $(0.06)$      \\
Bing use     & $0.11$        & $0.04$        & $0.05$        & $0.13^{*}$    & $0.04$       & $-0.03$       \\
                 & $(0.06)$      & $(0.06)$      & $(0.05)$      & $(0.06)$      & $(0.05)$     & $(0.05)$      \\
Ecosia use   & $-0.18$       & $-0.25$       & $-0.10$       & $-0.21$       & $-0.05$      & $0.12$        \\
                 & $(0.12)$      & $(0.13)$      & $(0.14)$      & $(0.13)$      & $(0.13)$     & $(0.13)$      \\
1|2              & $-0.83$       & $-0.06$       & $-0.19$       & $-1.36$       & $-0.96$      & $0.43$        \\
                 & $(0.94)$      & $(0.87)$      & $(0.87)$      & $(0.93)$      & $(0.85)$     & $(0.85)$      \\
2|3              & $-0.11$       & $0.70$        & $0.61$        & $-0.58$       & $-0.06$      & $1.23$        \\
                 & $(0.93)$      & $(0.87)$      & $(0.87)$      & $(0.91)$      & $(0.85)$     & $(0.85)$      \\
3|4              & $0.52$        & $1.30$        & $1.11$        & $-0.25$       & $0.53$       & $1.78^{*}$    \\
                 & $(0.92)$      & $(0.87)$      & $(0.87)$      & $(0.91)$      & $(0.85)$     & $(0.85)$      \\
4|5              & $1.27$        & $2.15^{*}$    & $1.69$        & $0.66$        & $1.27$       & $2.53^{**}$   \\
                 & $(0.92)$      & $(0.88)$      & $(0.87)$      & $(0.91)$      & $(0.85)$     & $(0.86)$      \\
5|6              & $2.01^{*}$    & $2.89^{**}$   & $2.32^{**}$   & $1.28$        & $1.91^{*}$   & $3.01^{***}$  \\
                 & $(0.92)$      & $(0.88)$      & $(0.87)$      & $(0.91)$      & $(0.85)$     & $(0.86)$      \\
6|7              & $3.24^{***}$  & $3.82^{***}$  & $3.07^{***}$  & $2.58^{**}$   & $2.99^{***}$ & $3.70^{***}$  \\
                 & $(0.93)$      & $(0.88)$      & $(0.88)$      & $(0.91)$      & $(0.86)$     & $(0.87)$      \\
\hline
AIC              & $1044.04$     & $1342.68$     & $1386.32$     & $1046.94$     & $1444.95$    & $1446.63$     \\
BIC              & $1123.31$     & $1421.95$     & $1465.59$     & $1126.22$     & $1524.22$    & $1525.91$     \\
Log Likelihood   & $-502.02$     & $-651.34$     & $-673.16$     & $-503.47$     & $-702.47$    & $-703.32$     \\
Deviance         & $1004.04$     & $1302.68$     & $1346.32$     & $1006.94$     & $1404.95$    & $1406.63$     \\
Num. obs.        & $389$         & $389$         & $389$         & $389$         & $389$        & $389$         \\
\hline
\multicolumn{7}{l}{\scriptsize{$^{***}p<0.001$; $^{**}p<0.01$; $^{*}p<0.05$}}
\end{tabular}
\caption{Additional regression models (using gender (G) rather than sex as one of the independent variables).}
\label{table:coefficientsapp}
\end{center}
\end{table}

\begin{table}
\begin{center}
\begin{tabular}{l c c c}
\hline
 & Mis: Inform & Mis: Reduce & Mis: Remove \\
\hline
Age              & $0.00$        & $-0.01$       & $0.01$        \\
                 & $(0.01)$      & $(0.01)$      & $(0.01)$      \\
Sex (Male)          & $0.04$        & $-0.23$       & $-0.16$       \\
                 & $(0.21)$      & $(0.19)$      & $(0.19)$      \\
Education       & $-0.06$       & $0.01$        & $-0.04$       \\
                 & $(0.08)$      & $(0.07)$      & $(0.07)$      \\
Ethnicity (White)  & $0.40$        & $0.01$        & $-0.14$       \\
                 & $(0.24)$      & $(0.22)$      & $(0.22)$      \\
Trust in SE           & $0.57^{***}$  & $0.36^{**}$   & $0.19$        \\
                 & $(0.12)$      & $(0.12)$      & $(0.12)$      \\
SE Independence            & $0.08$        & $0.21^{**}$   & $0.26^{***}$  \\
                 & $(0.08)$      & $(0.07)$      & $(0.07)$      \\
Political ideology & $-0.25^{***}$ & $-0.21^{***}$ & $-0.21^{***}$ \\
                 & $(0.04)$      & $(0.04)$      & $(0.04)$      \\
DDG Use     & $-0.07$       & $-0.12^{*}$   & $-0.15^{**}$  \\
                 & $(0.06)$      & $(0.05)$      & $(0.05)$      \\
Yandex Use   & $-0.13$       & $0.01$        & $-0.02$       \\
                 & $(0.14)$      & $(0.14)$      & $(0.15)$      \\
Yahoo Use    & $0.02$        & $0.07$        & $0.08$        \\
                 & $(0.07)$      & $(0.06)$      & $(0.06)$      \\
Bing Use     & $0.10$        & $0.03$        & $0.04$        \\
                 & $(0.06)$      & $(0.06)$      & $(0.05)$      \\
Ecosia Use   & $-0.15$       & $-0.20$       & $-0.06$       \\
                 & $(0.12)$      & $(0.13)$      & $(0.14)$      \\
1|2              & $-1.74^{*}$   & $-1.76^{*}$   & $-1.35$       \\
                 & $(0.80)$      & $(0.74)$      & $(0.73)$      \\
2|3              & $-1.03$       & $-1.03$       & $-0.56$       \\
                 & $(0.79)$      & $(0.74)$      & $(0.73)$      \\
3|4              & $-0.41$       & $-0.44$       & $-0.06$       \\
                 & $(0.78)$      & $(0.73)$      & $(0.73)$      \\
4|5              & $0.33$        & $0.40$        & $0.52$        \\
                 & $(0.77)$      & $(0.73)$      & $(0.73)$      \\
5|6              & $1.07$        & $1.12$        & $1.14$        \\
                 & $(0.77)$      & $(0.73)$      & $(0.73)$      \\
6|7              & $2.30^{**}$   & $2.04^{**}$   & $1.88^{**}$   \\
                 & $(0.78)$      & $(0.73)$      & $(0.73)$      \\
\hline
AIC              & $1044.99$     & $1350.49$     & $1387.76$     \\
BIC              & $1116.33$     & $1421.83$     & $1459.10$     \\
Log Likelihood   & $-504.49$     & $-657.25$     & $-675.88$     \\
Deviance         & $1008.99$     & $1314.49$     & $1351.76$     \\
Num. obs.        & $389$         & $389$         & $389$         \\
\hline
\multicolumn{4}{l}{\scriptsize{$^{***}p<0.001$; $^{**}p<0.01$; $^{*}p<0.05$}}
\end{tabular}
\caption{Statistical models with Google Use omitted}
\label{table:coefficients_noGoogle}
\end{center}
\end{table}

\end{document}